\documentclass[conference]{IEEEtran}
\usepackage[sort,comma,numbers]{natbib}
\usepackage{graphicx}
\usepackage{subfig}
\usepackage{amsmath}
\usepackage{listings}
\usepackage{tikz}
\usepackage{multirow}
\usepackage{url}

\PassOptionsToPackage{hyphens}{url}\usepackage{hyperref}

\def\checkmark{\tikz\fill[scale=0.3](0,.35) -- (.25,0) -- (1,.7) -- (.25,.15) -- cycle;} 

\newcommand\Attack{\emph{MemJam}}

\newcommand{\tA}{$\mathcal A$}
\newcommand{\tB}{$\mathcal B$}

\AtBeginDocument{}

\title{\Attack: A False Dependency Attack against Constant-Time Crypto Implementations}

\author{\IEEEauthorblockN{Ahmad Moghimi}
\IEEEauthorblockA{Worcester Polytechnic Institute\\
amoghimi@wpi.edu}
\and
\IEEEauthorblockN{Thomas Eisenbarth}
\IEEEauthorblockA{Worcester Polytechnic Institute\\
teisenbarth@wpi.edu}
\and
\IEEEauthorblockN{Berk Sunar}
\IEEEauthorblockA{Worcester Polytechnic Institute\\
sunar@wpi.edu}}

\date{\today}

\begin{document}
\maketitle
\begin{abstract}
Cache attacks exploit memory access patterns of cryptographic implementations. Constant-Time implementation techniques have become an indispensable tool in fighting cache timing attacks. These techniques engineer the memory accesses of cryptographic operations to follow a uniform key independent pattern. However, the constant-time behavior is dependent on the underlying architecture, which can be highly complex and often incorporates unpublished features. \emph{CacheBleed} attack targets cache bank conflicts and thereby invalidates the assumption that microarchitectural side-channel adversaries can only observe memory with cache line granularity. In this work, we propose \Attack, a side-channel attack that exploits false dependency of memory read-after-write and provides a high quality intra cache level timing channel. As a proof of concept, we demonstrate the first key recovery attacks on a constant-time implementation of AES, and a SM4 implementation with cache protection in the current Intel Integrated Performance Primitives (Intel IPP) cryptographic library. Further, we demonstrate the first intra cache level timing attack on SGX by reproducing the AES key recovery results on an enclave that performs encryption using the aforementioned  constant-time implementation of AES. Our results show that we can not only use this side channel to efficiently attack memory dependent cryptographic operations but also to bypass proposed protections. Compared to \emph{CacheBleed}, which is limited to older processor generations, \Attack~is the first intra cache level attack applicable to all major Intel processors including the latest generations that support the SGX extension. 

\end{abstract}

\section{Introduction}

In cryptographic implementations, timing channels can be introduced by key dependent operations, which can be exploited by local or remote adversaries~\cite{brumley2005remote,osvik2006cache}. Modern microarchitectures are complex and support various shared resources, and the operating system (OS) maximizes the resource sharing among concurrent tasks~\cite{schimmel1994unix,marr2002hyper}. From a security standpoint, concurrent tasks with different permissions share the same hardware resources, and these resources can expose exploitable timing channels. A typical model for exploiting microarchitectural timing channels is for a spy process to cause resource contention with a victim process and to measure the timing of its own or of the victim operations~\cite{ristenpart2009hey,aciiccmez2007new,tromer2010efficient,irazoqui2015s}. The observed timing behavior give adversaries strong evidence on the victim's resource usage pattern, thus they leak critical runtime data. Among the shared resources, attacks on cache have received significant attention, and their practicality have been demonstrated in scenarios such as cloud computing~\cite{ristenpart2009hey,zhang2012cross,irazoqui2015s,inci2016cache,yarom2014flush,gruss2016flush}. A distinguishable feature of cache attacks is the ability to track memory accesses with high temporal and spatial resolution. Thus, they excel at exploiting cryptographic implementations with secret dependent memory accesses~\cite{tsunoo2003cryptanalysis,osvik2006cache,benger2014ooh,inci2015seriously}. Examples of such vulnerable implementations include using S-Box tables~\cite{webster1986design}, and efficient implementations of modular exponentiation~\cite{kocc1995analysis}. 

The weakness of key dependent cache activities has motivated researchers and practitioners to protect cryptographic implementations against cache attacks~\cite{brickell2006software,tromer2010efficient}. The simplest approach is to minimize the memory footprint of lookup tables. Using a single 8-Bit S-Box in Advanced Encryption Standard (AES) rather than T-Tables makes cache attacks on AES inefficient in a noisy environment, since the adversary can only distinguish accesses between 4 different cache lines. Combining small tables with cache state normalization, i.e., loading all table entries into cache before each operation, defeats cache attacks in asynchronous mode, where the adversary is only able to perform one observation per operation. More advanced side channels such as exploitation of the thread scheduler~\cite{gullasch2011cache}, cache attack on interrupted execution of Intel Software Guard eXtension (SGX) ~\cite{moghimi2017cachezoom}, performance degradation~\cite{allan2016amplifying} and leakage of other microarchitectural resources~\cite{aciiccmez2007predicting,aciiccmez2010new} remind us the importance of constant-time software implementations. One way to achieve constant-time memory behavior, is the adoption of small tables in combination with accessing all cache lines on each lookup~\cite{tromer2010efficient}. The overhead would be limited and is minimized by the parallelism we can achieve in modern processors. Another constant-time approach adopted by some public cryptographic schemes is interleaving the multipliers in memory known as scatter-gather technique~\cite{brickell2006mitigating}.  

Constant-time implementations have effectively eliminated the first generation of timing attacks that exploit obvious key dependent leakages. The common view is that performance penalty is the only downside which, once paid, there is no need to be further worried. However, this is far from the reality and constant-time implementations may actually give a false sense of security. A commonly overlooked fact is that constant-time implementations and related protections are relative to the underlying hardware~\cite{ge2016contemporary}. In fact, there are major obstacles preventing us from obtaining true constant-time behavior. Processors constantly evolve with new microarchitectural features rolled quietly with each new release and the variety of such subtle features makes comprehensive evaluation impossible. A great example is the cache bank conflicts attack on OpenSSL RSA scatter-gather implementation: it shows that adversaries with intra cache level resolution can successfully bypass constant-time techniques relied on cache-line granularity \cite{yarom2017cachebleed}. As a consequence, what might appear as a perfect constant-time implementation becomes insecure in the next processor release--or worse--an unrecognized behavior might be discovered, invalidating the earlier assumption.

\subsection{Our Contribution}
We propose an attack named \Attack~by exploiting false dependency of memory read-after-write, and demonstrate key recovery against two different cryptographic implementations which are secure against cache attacks with experimental results on both regular and SGX environments. In summary: 
\begin{itemize}

\item \textbf{False Dependency Attack:} A side-channel attack on the false dependency of memory read-after-write. We show how to dramatically slow down the victim's accesses to specific memory blocks, and how this read latency can be exploited to recover low address bits of the victim's memory accesses.

\item \textbf{Attack on protected AES and SM4:} Attacks utilizing the intra cache level information on AES and SM4 implementations protected against cache attacks. The implementations are chosen from Intel Integrated Performance Primitives (Intel IPP), which is optimized for both security and speed.

\item \textbf{Attack on SGX Enclave:} The first intra cache level attack against SGX Enclaves supported by key recovery results on the contant-time AES implementation. The aforementioned constant-time implementation of AES is part of the SGX SDK source code.

\item \textbf{Protection Bypass:} Bypasses of remarkable protections such as proposals based on constant-time techniques~\cite{brickell2006mitigating,tromer2010efficient}, static and runtime analysis~\cite{zhang2016cloudradar,irazoqui2016mascat} and cache architecture~\cite{liu2016catalyst,costan2016sanctum,kayaalp2017ric,xu2017vcat}.
\end{itemize}

\subsection{Experimental Setup and Generic Assumptions}
Our experimental setup is a Dell XPS 8920 desktop machine with Intel(R) Core i7-7700 processor running Ubuntu 16.04. The Core i7-7700 has 4 hyper-threaded physical cores. Our only assumptions are that the attacker is able to co-locate on one of the logical processor pairs within the same physical core as the victim. In the cryptographic attacks, the attacker can measure the time of victim encryption. The attacker further knows which cryptographic implementation is used by the victim, but she does not need to have any knowledge of the victim's binary or the offset of the S-Box tables. We will discuss assumptions that are specific to the attack on SGX at Section~\ref{sec:sgx}.

\section{Related Work}
\textbf{Side channels} including power, electromagnetic and timing channels have been studied for a few decades~\cite{kocher2011introduction,brumley2005remote,carluccio2005electromagnetic}. Timing side channels can be constructed through the processor cache to perform key recovery attacks against cryptographic operations such as  RSA~\cite{inci2015seriously}, ECDSA~\cite{benger2014ooh}, ElGamal~\cite{zhang2012cross}, DES~\cite{tsunoo2003cryptanalysis} and AES~\cite{irazoqui2015s,osvik2006cache}. On multiprocessor systems, attacks on the shared LLC---a shared resource among all the cores---perform well even when attacker and victim reside in different cores~\cite{irazoqui2015s}. Flush+Reload, Prime+Probe, Evict+Reload, and Flush+Flush are some of the proposed attack methodologies with different adversarial scenarios~\cite{yarom2014flush,gruss2016flush,osvik2006cache}. Performance degradation attacks can improve the channel resolution~\cite{gullasch2011cache,allan2016amplifying}. LLC attacks are highly practical in cloud, where an attacker can identify where a particular victim is located~\cite{ristenpart2009hey,zhang2012cross}. Despite the applicability of LLC attacks, attacks on core-private resources such as L1 cache are as important~\cite{aciiccmez2010new,bonneau2006cache}. Attacks on SGX in a system level adversarial scenario are notable examples~\cite{lee2016inferring,moghimi2017cachezoom}. There are other shared resources, which can be utilized to construct timing channels~\cite{ge2016survey}. Exploitation of Branch Target Buffer (BTB) leaks if a branch has been taken by a victim process~\cite{aciiccmez2007predicting,aciiccmez2010new,lee2016inferring}. Logical units within the processor can leak information about the arithmetic operations~\cite{aciicmez2007cheap,andrysco2015subnormal}. \emph{CacheBleed} proposes cache bank conflicts and false dependency of memory write-after-read as side channels with intra-cache granularity~\cite{yarom2017cachebleed}. However, cache bank conflicts leakage does not exist on current Intel processors, and we verify the authors' claim that the proposed write-after-read false dependency side channel does not allow efficient attacks. 

\textbf{Defense:} software and hardware strategies have been proposed such as alternative lookup tables, data-independent memory access pattern, static or disabled cache, and cache state normalization to defend against cache attacks~\cite{tromer2010efficient}. Scatter-Gather techniques have been adopted by RSA and ECC implementations~\cite{brickell2006mitigating}. In particular, introducing redundancy and randomness to the S-Box tables for AES has been proposed~\cite{brickell2006software}. A custom memory manager~\cite{zhou2016software}, relaxed inclusion caches~\cite{kayaalp2017ric} and solutions based on cache allocation technology (CAT) such as Catalyst~\cite{liu2016catalyst} and vCat~\cite{xu2017vcat} are proposed to defend against LLC contention. Sanctum~\cite{costan2016sanctum} and Ozone~\cite{aweke2017ozone} are new processor designs with respect to cache attacks. Detection-based countermeasures have also been proposed using performance counters, which can be used to detect cache attacks in cloud environments~\cite{zhang2016cloudradar,briongos2017cacheshield}. MASCAT~\cite{irazoqui2016mascat} is proposed to block cache attacks with code analysis techniques. CachD~\cite{CachD} detects potential cache leakage in the production software. Nonetheless, these proposals assume that the adversary cannot distinguish accesses within a cache line. That is, attacks with intra cache-line granularity are considered out-of-scope. Doychev et al. proposed the only software leakage detector that consider full address bits as its leakage model~\cite{doychev2017rigorous}. 

\section{Background}
\subsection{Multitasking}
The memory management subsystem shares the dynamic random-access memory (DRAM) among all concurrent tasks, in which a virtual memory region is allocated for each task transparent to the physical memory. Each task is able to use its entire virtual address space without meddling of memory accesses from others. Memory allocations are performed in pages, which each virtual memory page can be stored in a DRAM page with a virtual-to-physical page mapping. The logical processors are also shared among these tasks and each logical processor executes instructions from one task at a time, and switches to another task. Memory write and read instructions work with virtual addresses, and the virtual address is translated to the corresponding physical address to perform the memory operation. The OS is responsible for page directory management and virtual page allocation. The OS assists the processor to perform virtual-to-physical address translation by performing an expensive page walk. The processor saves the address translation results in a memory known as Translation Look-aside Buffer (TLB) to avoid the software overhead introduced by the OS. Intel microarchitecture follows a multi-stage pipeline and adopts different optimization techniques to maximize the parallelism and multitasking during the pipeline stages~\cite{inteloptimze}. Among these techniques, hyper-threading allows each core to run multiple concurrent threads, and each thread shares all the core-private resources. As a result, if one resource is busy by a thread, other threads can consume the remaining available resources. Hyper-threading is abstracted to the software stack: OS and applications interact with the logical processors.

\subsection{Cache Memory}
DRAM memory is slow compared to the internal CPU components. Modern microarchitectures take advantage of a hierarchy of cache memories to fill the speed gap. Intel processors have two levels of core-private cache (L1, L2), and a Last Level Cache (LLC) shared among all cores. The closer the cache memory is to the processor, the faster, but also smaller it is compared to the next level cache. Cache memory is organized into different sets, and each set can store some number of cache lines. The cache line size, which is 64 byte, is the block size for all memory operations outside of the CPU. The higher bits of the physical address of each cache line is used to determine which set to store/load the cache line. When the processor tries to access a cache line, a cache hit or miss occurs respective of its existence in the relevant cache set. If a cache miss occurs, the memory line will be stored to all 3 levels of cache and to the determined sets. Reloads from the same address would be much faster when the memory line exists in cache. In a multicore system, the processor has to keep cache consistent among all levels. In Intel architecture, cache lines follow a write-back policy, i.e, if the data in L1 cache is overwritten, all other levels will be updated. The LLC is inclusive of L2 and L1 caches, which means that if a cache line in LLC is evicted, the corresponding L1 and L2 cache lines will also be evicted~\cite{inteloptimze}. These policies help to avoid stale cached data where one processor reads invalid data mutated by another processor. 

\subsection{L1 Cache Bottlenecks}
L1 cache port has a limited bandwidth and simultaneous accesses will be block each other. This bottleneck is critical in super-scalar multiprocessor systems. Older processors' generation adopted multiple banks as a workaround to this problem~\cite{agneroptimize}, in which each bank can operate independently and serve one request at a time. While this partially solved the bandwidth limit, it creates the cache bank conflicts phenomena which simultaneous accesses to the same bank will be blocked. Intel resolved the cache bank conflicts issue with the Haswell generation~\cite{inteloptimze}. Another bottleneck mentioned in various resources is due to the false dependency of memory addresses with the same cache set and offset~\cite{inteloptimze,agneroptimize}. Simultaneous read and write with addresses that are multiples of 4\,kB is not possible, and they halt each other. The processor cannot determine the dependency from the virtual address, and addresses with the same last 12 bits have the chance to map to the same physical address. Such simultaneous access can happen between two logical processors and/or during the out-of-order execution, where there is a chance that a memory write/read might be dependent on a memory read/write with the same last 12 bits of address. Such dependencies cannot be determined on the fly, thus they cause latency.

\subsection{Cache Attacks}
Cache attacks can be exploited by adversaries where they share system cache memory with benign users. In scenarios where the adversary can colocate with a victim on the same core, she can attack core-private resources such as L1 cache, e.g., OS adversaries~\cite{lee2016inferring,moghimi2017cachezoom}. In cloud environment, virtualization platforms allow sharing of logical processors to different VMs; however, attacks on the shared LLC have a higher impact, since LLC is shared across all the cores. In cache timing attacks, the attacker either measure the timing of the victim operations, e.g, \emph{Evict+Time}~\cite{osvik2006cache} or the timing of his own memory accesses, e.g, \emph{Prime+Probe}~\cite{irazoqui2015s}. The attacker needs to have access to an accurate time resource such as the \emph{RDTSC} instruction. In the basic form, attacks are performed by one observation per entire operation. In certain scenarios, these attacks can be improved by interrupting the victim and collecting information about the intermediate memory states. Side channel attacks exploiting cache bank conflicts rely on synchronous resource contention. \emph{CacheBleed} methodology is somewhat similar to Prime+Probe, where the attacker performs repeated operations, and measures it's own access time~\cite{yarom2017cachebleed}. In a cache bank conflicts attack, the adversary repeatedly performs simultaneous reads to the same cache bank and measures their completion time. A victim on a colocated logical processor who access the same cache bank would cause latency to the attacker's memory reads.

\begin{figure*}[t!]
\centering
\begin{minipage}{.65\textwidth}
\begin{minipage}{.46\textwidth}
\begin{lstlisting}[label={lst:probe_read},caption=Probe Reads,captionpos=b, frame=tb, basicstyle=\small]
loop:
rdtscp;
mov %eax, (%r9);
movb 0x0000(%r10), %al;
movb 0x1000(%r10), %al;
movb 0x2000(%r10), %al;
movb 0x3000(%r10), %al;
movb 0x4000(%r10), %al;
movb 0x5000(%r10), %al;
movb 0x6000(%r10), %al;
movb 0x7000(%r10), %al;
add $4, %r9;
dec %r11;
jnz loop;
\end{lstlisting}
\end{minipage}\hfil
\begin{minipage}{.46\textwidth}
\begin{lstlisting}[label={lst:probe_write},caption=Probe Writes,captionpos=b, frame=tb, basicstyle=\small]
loop:
rdtscp
mov %eax, (%r9);
movb %al, 0x0000(%r10);
movb %al, 0x1000(%r10);
movb %al, 0x2000(%r10);
movb %al, 0x3000(%r10);
movb %al, 0x4000(%r10);
movb %al, 0x5000(%r10);
movb %al, 0x6000(%r10);
movb %al, 0x7000(%r10);
add $4, %r9
dec %r11
jnz loop
\end{lstlisting}
\end{minipage}\hfil
\caption*{Listings 1 and 2 are used to probe 8 parallel reads and writes, respectively. \emph{r9} points to a measurement buffer, and \emph{r11} is initialized with the probe count.}
\end{minipage}\hfil
\begin{minipage}{.27\textwidth}
\includegraphics[width=.99\linewidth]{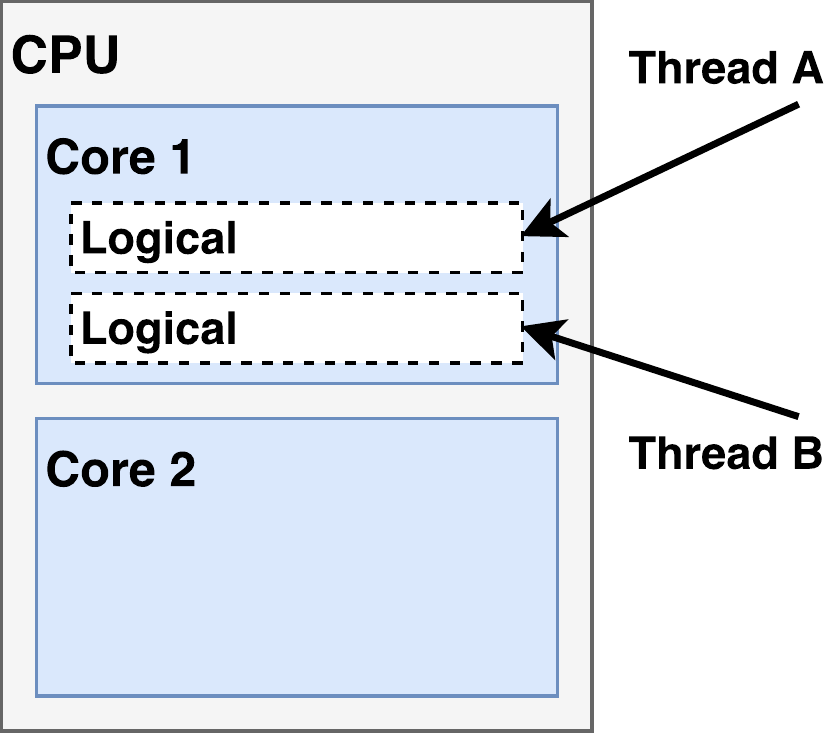}
\caption{Based on the attack model, thread \emph{A} and \emph{B} both run on the same core, and introduce and probe stall hazards.}
\label{fig:hypert}
\end{minipage}
\end{figure*}

\section{\Attack: Read-After-Write Attack}

\Attack\ utilizes \emph{false dependencies}. Data dependency occurs when an instruction refers to the data of a preceding instruction. In pipelined designs, hazards and pipeline stalls can occur from dependencies if the previous instruction has not finished. There are cases where false dependencies occur, i.e. the pipeline stalls even though there is no true dependency. Reasons for false dependencies are register reuse and limited address space for the Arithmetic Logic Unit (ALU). False dependencies degrade instruction level parallelism and cause overhead. The processor eliminates false dependencies arising from register reuse by a register renaming approach. However, there exist other false dependencies that need to be addressed during the software optimization~\cite{inteldev,inteloptimze}.

In this work, we focus on a critical false dependency mentioned as \emph{4K Aliasing} where data that is multiples of 4k apart in the address space is seen as dependent. 4k Aliasing happens due to virtual addressing of L1 cache, where data is accessed using virtual addresses, but tagged and stored using physical addresses. More than one virtual addresses can refer to the same data with the same physical address and the determination of dependency for concurrent memory accesses, requires virtual address translation. Physical and virtual address share the last 12 bits, and any data accesses whose addresses differ in the last 12 bits (i.e. the distance is not 4k) cannot have a dependency. For the fairly rare remaining cases, address translation needs to be done before resolving the dependency, which causes latency. Note that the \emph{granularity} of the potential dependency, i.e. whether two addresses are considered ``same'', depends also on the microarchitecture, as dependencies can occur at the \emph{word} or cache \emph{line} granularity (i.e.\ ignoring the last 2 or last 6 bits of the address, respectively). These rare false dependencies due to 4K aliasing can be exploited to attack memory, since the attacker can deliberately process falsely dependent data by matching the last 12 bits of his own address with a security critical data inside a victim process.

4K Aliasing has been mentioned in various places as an optimization problem existing on all major Intel processors~\cite{agneroptimize,inteloptimze}. We verify the results of Yarom et al.~\cite{yarom2017cachebleed}, the only security related work regarding false dependencies, which exploited \emph{write-after-read} dependencies. The resulting timing leakage by write stall after read is not sufficient to be used in any cryptographic attack. \Attack\ exploits a different channel due to the false dependency of~\emph{read-after-write}, which causes a higher latency and is thus simply observable. Intel Optimization Manual highlights the \emph{read-after-write} performance overhead in various sections~\cite{inteloptimze}. As described in Section~11.8, this hazard occurs when a memory write is closely followed by a read, and it causes the read to be reissued with a potential 5 cycles penalty\footnote{\texttt{LD\_BLOCKS\_PARTIAL.ADDRESS\_ALIAS} Performance Monitoring Unit (PMU) event counts the number of times reads were blocked.}. In Section B.1.4 on memory bounds, write operations are treated under the store bound category. In contrast to load bounds, Top-down Microarchitecture Analysis Method (TMAM)\footnote{Top-Down Characterization is a hierarchical organization of event-based metrics that identifies the dominant performance bottlenecks in an application.~\cite{tmam}} reports store bounds as fraction of cycles with low execution port utilization and small performance impact. These descriptions in various sections highlight that~\emph{read-after-write} stall is considered more critical than \emph{write-after-read} stall.


\subsection{Memory Dependency Fuzz Testing}
We performed a set of experiments to evaluate the memory dependency behavior between two logical processors. In these experiments, we have thread \tA\ and \tB\ running on the \emph{same} physical core, but on \emph{different} logical processors, as shown in Figure~\ref{fig:hypert}. Both threads perform memory operations; only thread \tB\ measures its timing and hence the timing impact of introduced false dependencies.

\begin{figure*}[!t]
\centering
\subfloat[RaR]{\includegraphics[width=.33\linewidth]{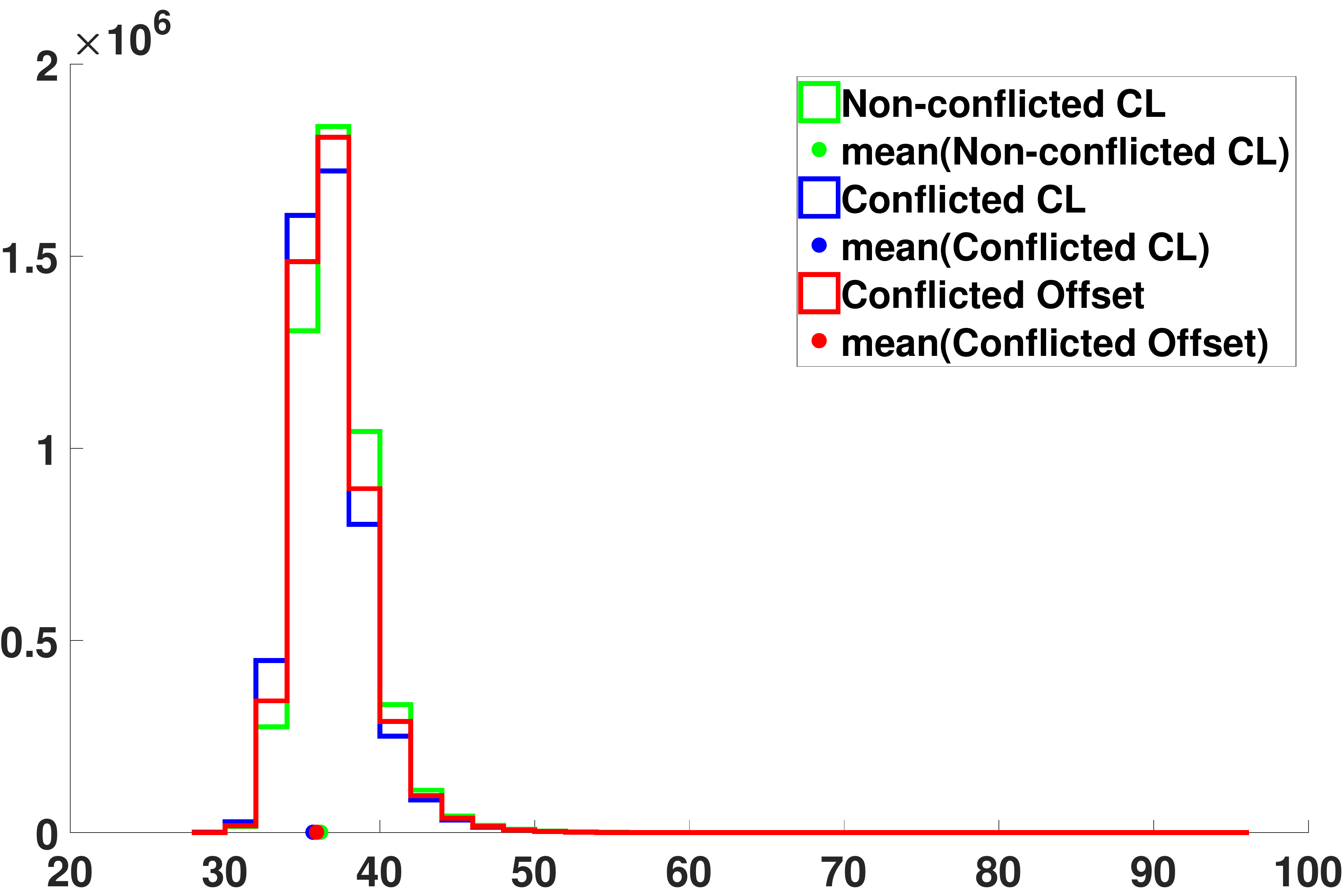}%
\label{fig:histogram_1}}
\hfil
\subfloat[WaR]{\includegraphics[width=.33\linewidth]{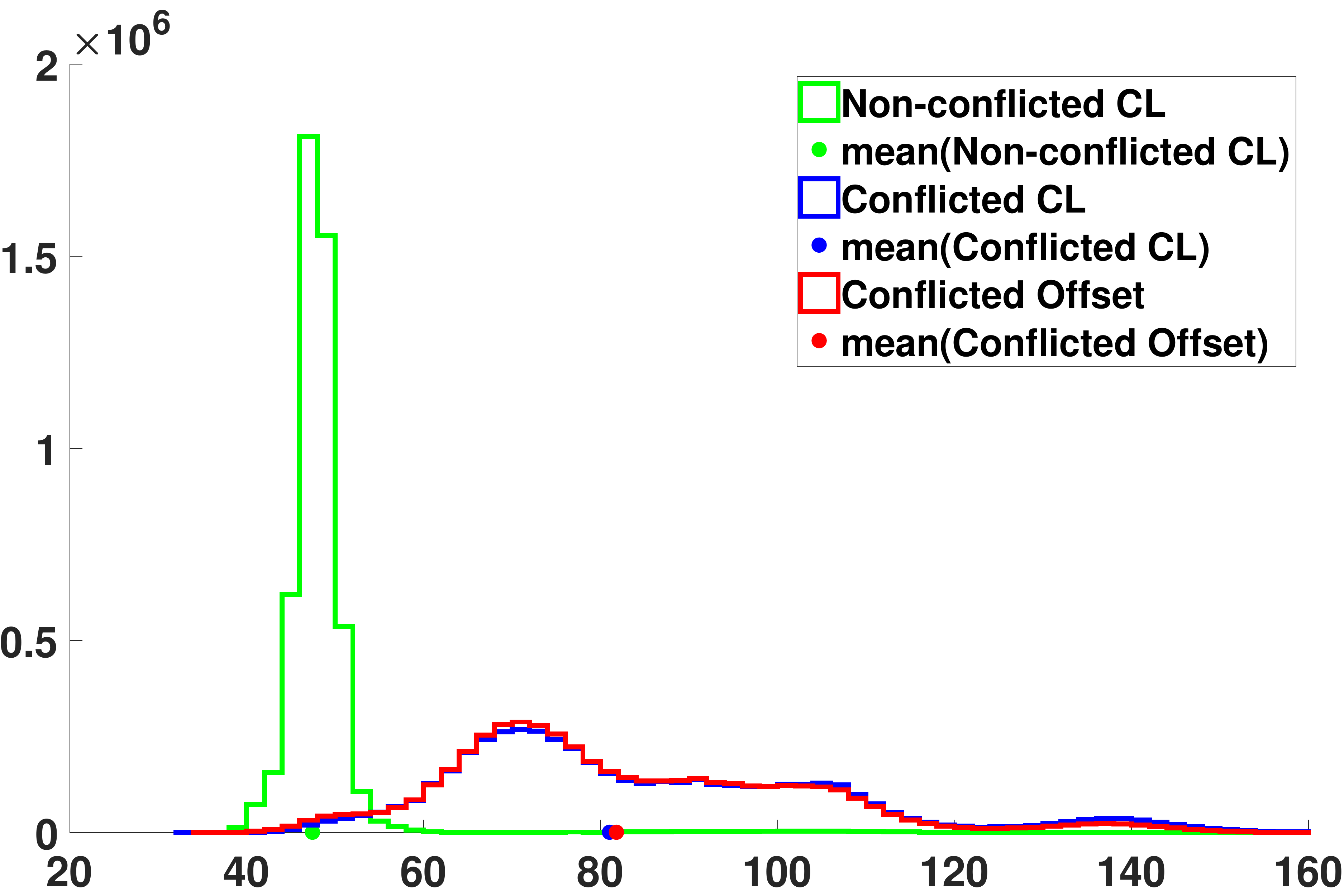}%
\label{fig:histogram_2}}
\hfil
\subfloat[RaW]{\includegraphics[width=.33\linewidth]{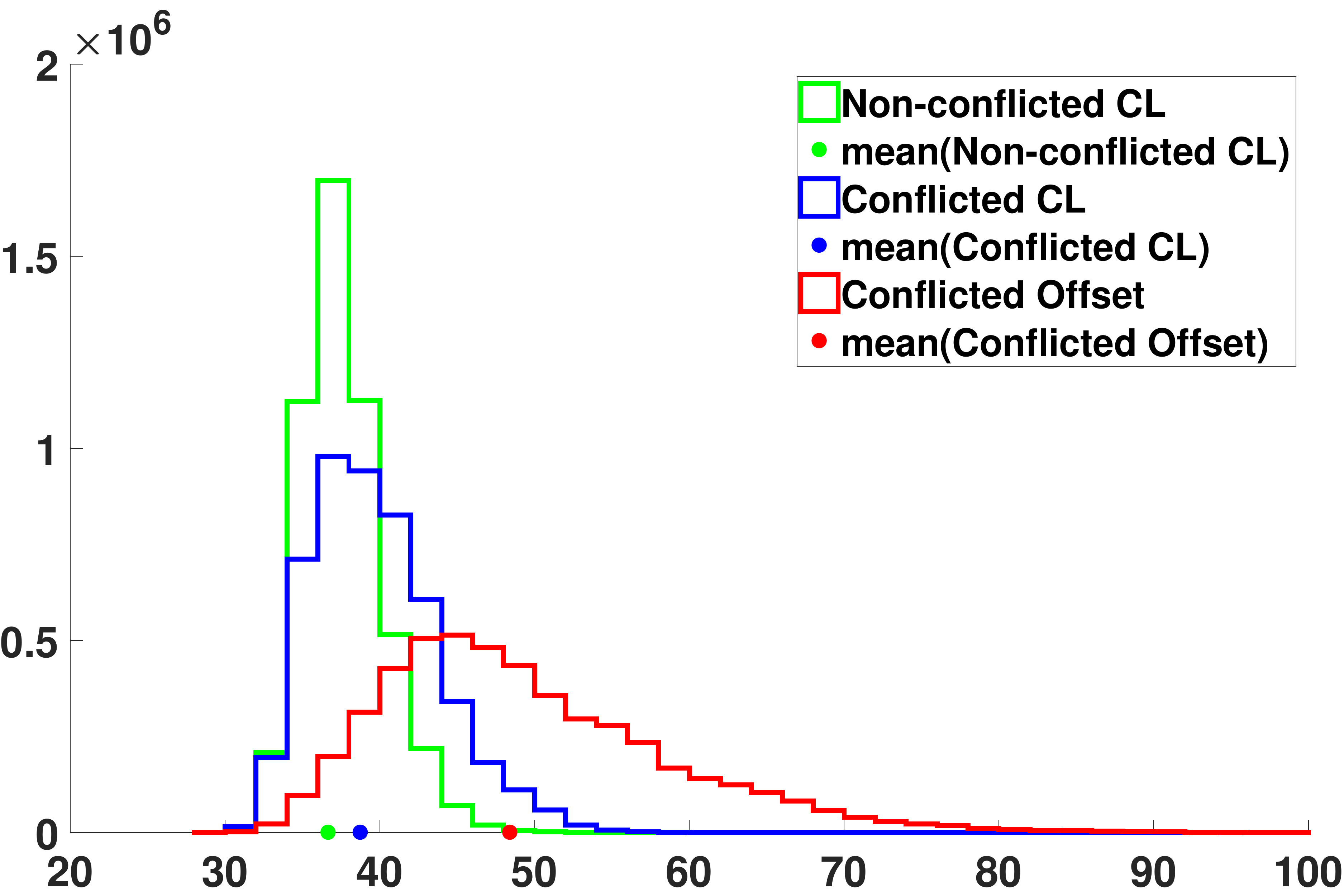}%
\label{fig:histogram_3}}
\caption{Three different scenario where different cache line (green), same cache line (blue) and same offset (red) have been accessed by two logical processors. Experiment $(c)$ on RaW latency has distinguishable characteristics for the conflicted word offset (red), while $(a)$ and $(b)$ feature nimble differences.}
\label{fig:histogram}
\end{figure*}

\smallskip
\noindent\textbf{Read-after-read (RaR):} 
In the first experiment, the two logical threads \tA\ and \tB\ read from the same shared cache and can potentially block each other. This experiment can reveal cache bank conflicts, as used by \emph{CacheBleed}~\cite{yarom2017cachebleed}.  
\tB\ uses Listing~\ref{lst:probe_read} to perform read measurements and \tA\ constantly reads from different memory offsets and tries to introduce conflicts. \tA\ reads from three different type of offsets: \textbf{(1)} Different cache line than \tB, \textbf{(2)} same cache line, but different offset than \tB , and \textbf{(3)} same cache line and same offset as \tB . As depicted, there is no obvious difference between the histograms for three cases in Figure~\ref{fig:histogram_1} verifying the lack of cache bank conflicts on 7th generation CPUs. 

\smallskip
\noindent\textbf{Write-after-read (WaR):} 
The histogram results for the second experiment on false dependency of write-after-read is shown in Figure~\ref{fig:histogram_2}, in which the cache line granularity is obvious. Thread \tA\ constantly reads from different type of memory offsets, while thread \tB\ uses Listing~\ref{lst:probe_write} to perform write measurements. The standard deviation for conflicted cache line (blue) and conflicted offset (red) between thread \tA\ and \tB\ is distinguishable from the green bar where there is no cache line conflict. This shows a high capacity cache granular behavior, but the slight difference between conflicted line and offset verifies the previous results stating a weak offset dependency~\cite{yarom2017cachebleed}.

\begin{figure}[!t]
\centering
\centering
\includegraphics[width=.98\linewidth]{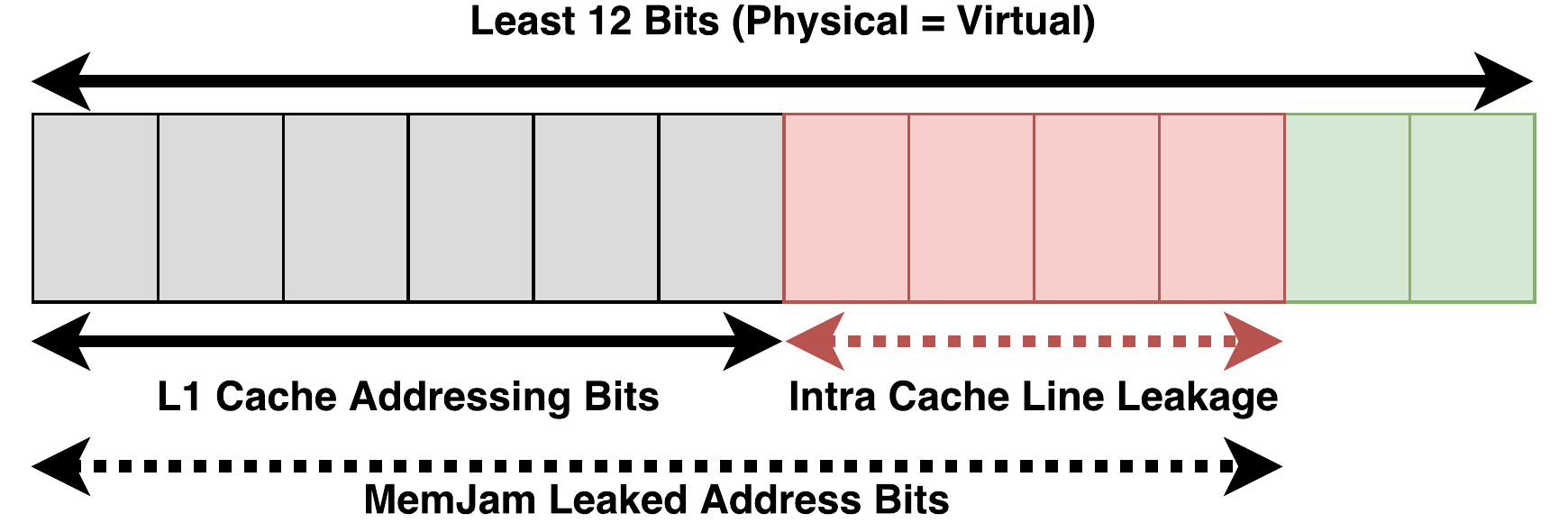}
\caption{Intra Cache Level Leakage:~\Attack~latency is related to 10 address bits, in which 4 bits are intra cache level bits.}
\label{fig:address}
\end{figure}

\smallskip
\noindent\textbf{Read-after-write (RaW):} 
Figure~\ref{fig:histogram_3} shows an experiment on measuring false dependency of read-after-write, in which, thread \tA\ constantly writes to different memory offsets. Thread \tB\ uses Listing~\ref{lst:probe_read} to perform read measurements. Accesses to three different types of offsets are clearly distinguishable. The conflicted cache line accesses (blue) are distinguishable from non-conflicted accesses (green). More importantly, conflicted accesses to the same offset (red) are also distinguishable from conflicted cache line accesses, resulting in a side channel with intra cache-line granularity. There is an average of 2 cycle penalty if the same cache line has been accessed, and a 10 cycle penalty if the same offset has been accessed. Note that the word offsets in our platform have 4 byte granularity. From an adversarial standpoint, this means that an adversary learns about bits 2-11 of the victim memory access, in which 4 bits (bits 2-5) are related to intra cache-line resolution, and thus goes beyond any other microarchitectural side channels known to exist on 6th and 7th generation Intel processors (Figure~\ref{fig:address}).

\begin{figure}[t!]
\centering
\includegraphics[width=.85\linewidth]{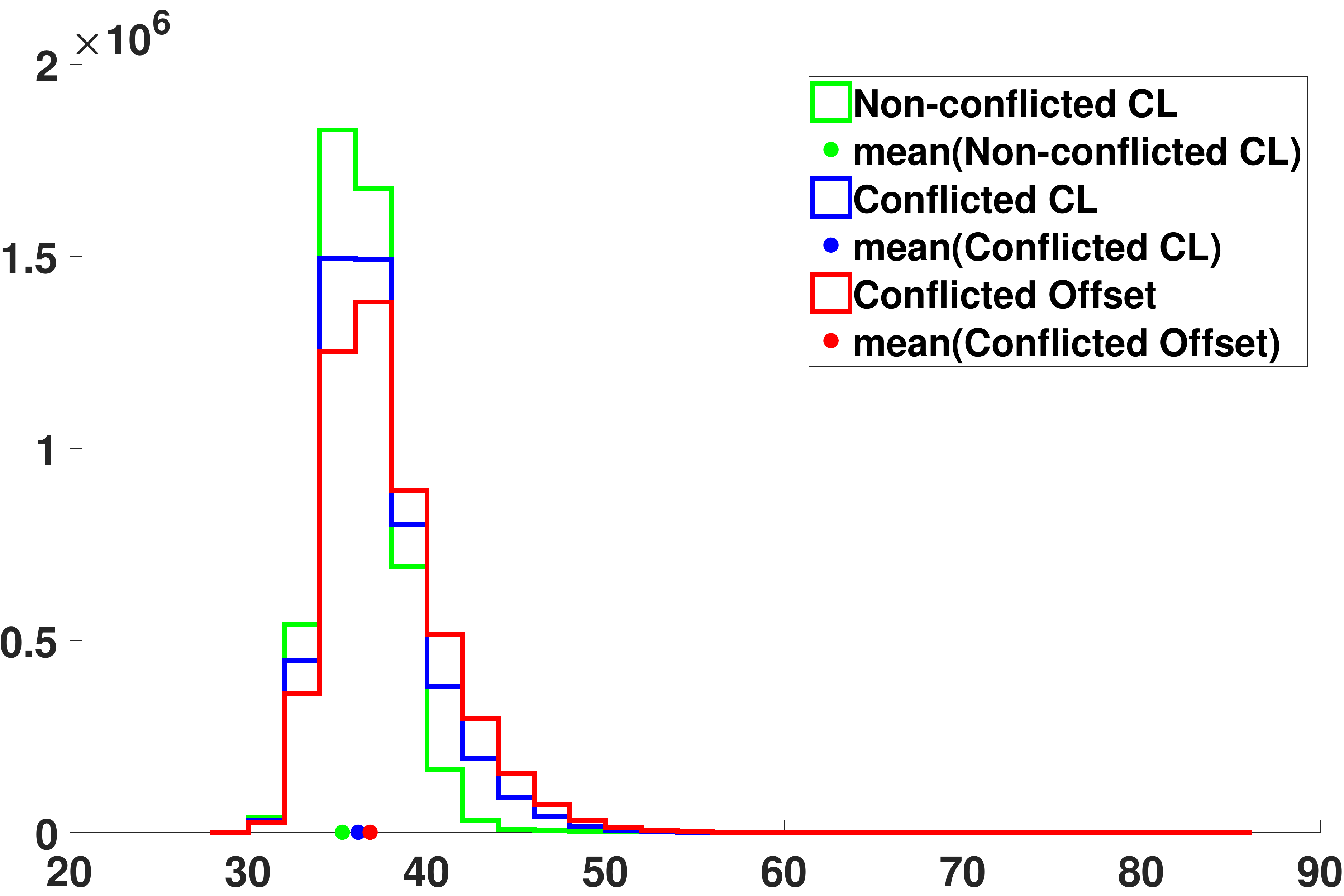}
\captionof{figure}{RawW: Compared to figure~\ref{fig:histogram_3}, this shows a lower impact on access latency.} 
\label{fig:slowwrite}
\end{figure}

\begin{figure}[t!]
\centering
\includegraphics[width=.85\linewidth]{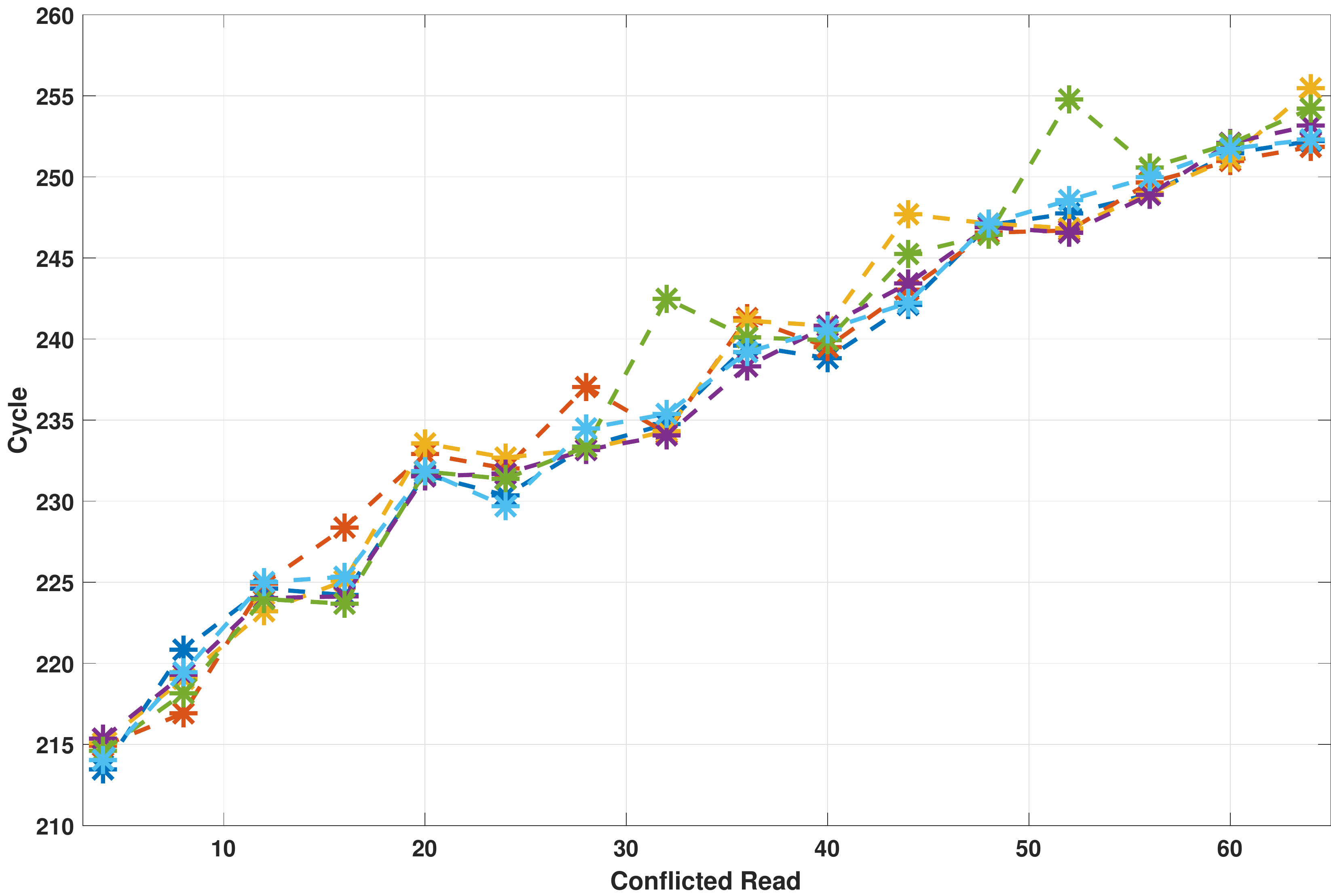}
\captionof{figure}{The cycle count for mixed operations with RaW conflicts. More conflicts cause higher delay.}
\label{fig:read_cost}
\end{figure}

\smallskip
\noindent\textbf{Read-after-weak-Write (RawW):} 
In this experiment on the read-after-write conflicts, we followed a less greedy strategy on the conflicting thread. Rather than constantly writing to the same offset, \tA\ executes write instructions to the same offset with some gaps filled with other memory accesses and instructions. As shown in Figure~\ref{fig:slowwrite}, the channel dramatically became less effective. This tells us that causing read access penalty will be more effective with constant writes to the same offset without additional instruction. In this regard, we will use Listing~\ref{lst:conflict} in our attack to achieve the maximum conflicts.

\begin{figure}[t!]
\centering
\begin{lstlisting}[label={lst:conflict},caption=Write Conflict Loop: Unnecessarily instructions are avoided to minimize usage of other processor units\, thus it maximizes the RaW conflict effect.,captionpos=b, frame=tb, basicstyle=\small]
mov %[target], %rax;
write_loop:
  .rept 100;
  movb $0, (%rax);
  .endr;
jmp write_loop;
\end{lstlisting}
\end{figure}

\smallskip
\noindent\textbf{Read-after-Write Latency:} 
In the last experiment, we tested the delay of execution over a varying number of conflicting reads. We created a code stub that includes 64 memory read instructions and a random combination of instructions between memory reads to create a more realistic computation. The combination is chosen in a way to avoid unexpected halts and to maintain the parallelism of all read operations. We measure the execution time of this computation on \tB, while \tA\ is writing to a conflicting offset. First, we measured the computation with 64 memory reads to addresses without conflicts. Our randomly generated code stub takes an average of 210 cycles to execute. On each step of the experiments, as shown in Figure~\ref{fig:read_cost}, we change some of the memory offsets to have the same last 12 bits of address as of \tA\'s conflicting write offset. We observe a growth on read accesses' latency by increasing the number of conflicting reads. Figure~\ref{fig:read_cost} shows the results for a number of experiments. In all of them, the overall execution time of \tB\ is strongly dependent on the number of conflicting reads. Hence, we can use the RaW dependency to introduce strong timing behavior using bits 2-11 of a chosen target memory address.

\section{\Attack~Correlation  Attack}

\Attack\ uses read-after-write false dependencies to introduce timing behavior to otherwise constant-time implementations. The resulting latency is then exploited using a correlation attack. \Attack\ proceeds with the following steps:
\begin{enumerate}
\item Attacker launches a process constantly writing to an address using Listing~\ref{lst:conflict} where the last 12 bits match the virtual memory offset of a \emph{critical} data that is read in the victim's process.
\item While the attacker's conflicting process is running, attacker queries the victim for encryption and records a ciphertext and execution time pair of the victim. Higher time infers more accesses to the \emph{critical} offset. 
\item Attacker repeats the previous step collecting ciphertext and time pairs. 
\end{enumerate}

\begin{figure*}[!t]
\centering
\subfloat{\includegraphics[width=.7\linewidth]{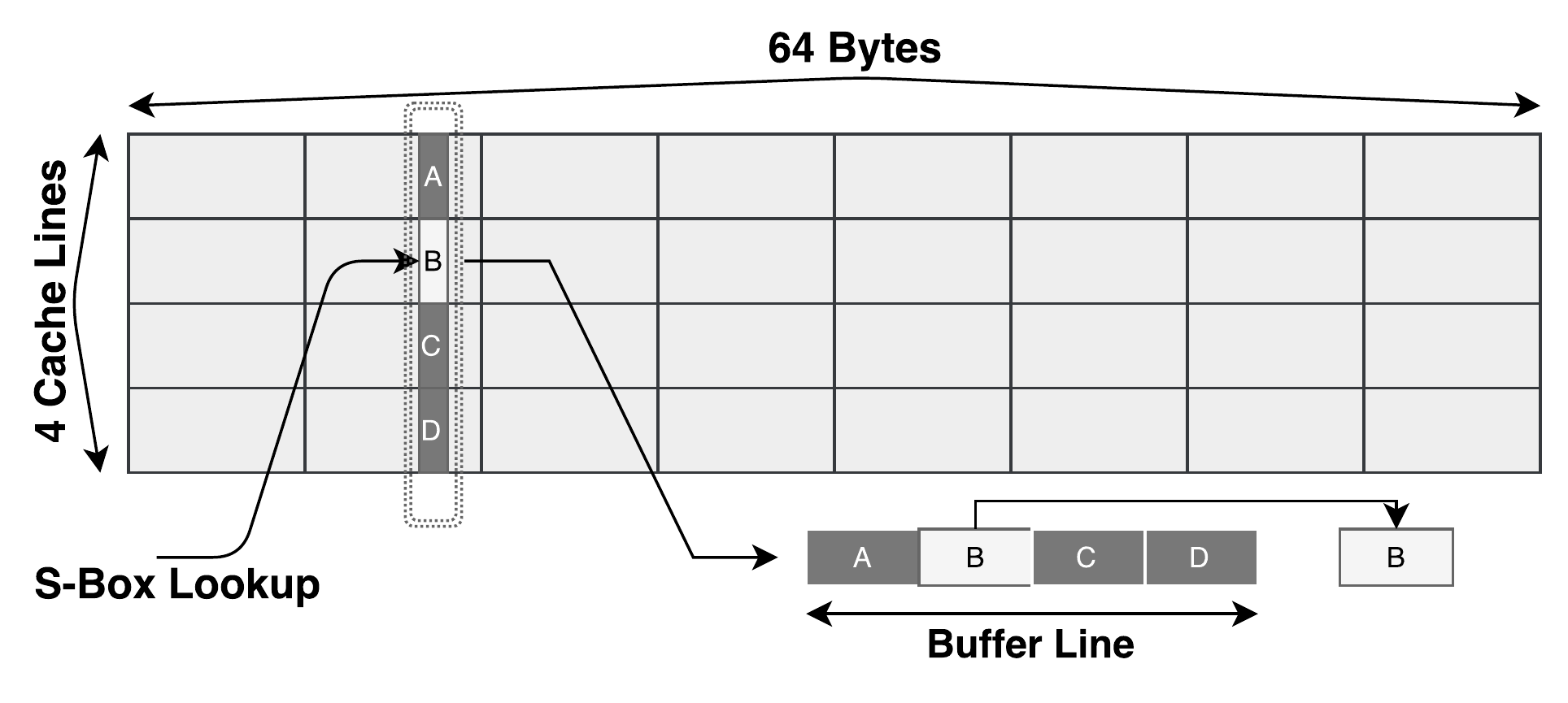}%
\label{l}}
\hfil
\caption{Constant-Time table lookup used by Intel IPP: Each lookup preloads 4 values to a cache aligned buffer, thus it accesses all the 4 S-Box cache lines. The actual output will be chosen from the buffer using the high address bits.}
\label{fig:aes_sbox}
\end{figure*}

The attack methodology resembles the \emph{Evict+Time} strategy originally proposed by Tromer et al.~\cite{tromer2010efficient}, except that the attacker uses false dependencies rather than evictions to slow down the target \emph{and} that the slowdown only applies to an 4-byte block of a cache line. Furthermore, \emph{all} of the victim's accesses addresses with the same last 12 bits are slowed down while an eviction only slows the first memory access(es).

Based on the intra cache level leakage in Figure~\ref{fig:address}, we divide a 64 byte cache line into 4-byte blocks and hypothesize that the access counts to a block are correlated with the running time of victim, while the attacker jams memory reads to that block, i.e, the attacker expects to observe a higher time when there are more accesses by the victim to the targeted 4-byte block and lower time when there are lower number of accesses. Based on this hypothesis, we apply a classical correlation based side-channel approach~\cite{kocher2011introduction} to attack implementations of two different block ciphers, namely AES and SM4, a standard cipher. SM4 among AES, Triple DES, and RC4 are the only available symmetric ciphers as part of Intel's IPP crypto library~\cite{intelIPP}\footnote{Patents investigated by Intel verify the importance of SM4~\cite{gueron2016sm4,wolrich2016sms4,yap2016sms4}}. Both implementations have optimizations to hinder cache attacks. In fact, the AES implementation features a constant cache profile and can thus be considered resistant to most microarchitectural attacks including cache attacks and high-resolution attacks as described in~\cite{moghimi2017cachezoom}. \Attack\ can still extract the keys from both implementations due to the intra cache-line spatial resolution as depicted in Figure~\ref{fig:address}. We describe the targeted implementations next, as well as the correlation models we use to attack them.

\subsection{Attack 1: IPP Constant-Time AES} \label{sec:aes}

AES is a cipher based on substitution permutation network (SPN) with 10 rounds supporting 128-bit blocks and 128/192/256-bit keys~\cite{daemen2013design}. The SubBytes is a security-critical operation and the straightforward way to implement AES SubBytes operation efficiently in software is to use lookup tables. SubBytes operates on each byte of cipher state, and it maps an 8-bit input to an 8-bit output using a non-linear function. A precomputed 256 byte lookup table known as S-Box can be used to avoid recomputation. There are efficient implementations using T-Tables that output 32-bit states and combine SubBytes and MixColumns operations. T-Table implementations are highly vulnerable to cache attacks. During AES rounds, a state table is initiated with the plaintext, and it holds the intermediate state of the cipher. Round keys are mixed with states, which are critical S-Box inputs and the main source of leakage. Hence, even an adversary who can partially determine which entry of the S-Box has been accessed is able to learn some information about the key.

Among the efforts to make AES implementations more secure against cache attacks, \texttt{Safe2Encrypt\_RIJ128} function from Intel IPP cryptographic library is noteworthy. This implementation is the only production-level AES software implementation that features true cache constant-time behavior and does not utilize hardware extensions such as AES-NI or SSSE3 instruction sets. This implementation is also part of the Linux SGX SDK~\cite{linuxsgx} and can be used for production code if the SDK is compiled from the scratch, i.e., it does not use prebuilt binaries. We verified the match between the implementation in Intel IPP binary and SGX SDK source code through reverse engineering. This implementation follows a very simple direction: \textbf{(1)} it implements AES using 256\,byte S-Box lookups without any optimization such as T-Tables, \textbf{(2)} instead of accessing a single byte of memory on each S-Box lookup, it fetches four values from the same vertical column of 4 different cache lines and saves them to a local cache aligned buffer, finally, \textbf{(3)} It performs the S-Box replacement by picking the correct S-Box entry from the local buffer. This implementation is depicted in Figure~\ref{fig:aes_sbox}. This implementation protects AES against any kind of cache attacks, as the attacker sees a constant cache access pattern: The S-Box table only occupies 4 cache lines, and on each SubBytes operation, all of them will sequentially be accessed. This implementation can be executed in less than 2000 cycles on a recent laptop processor. This is fast enough for many cryptographic applications, and it provides full protection against cache attacks, even if the attacker can interrupt the execution pipeline.

\begin{figure*}[t!]
\centering
\subfloat{\includegraphics[width=.9\linewidth]{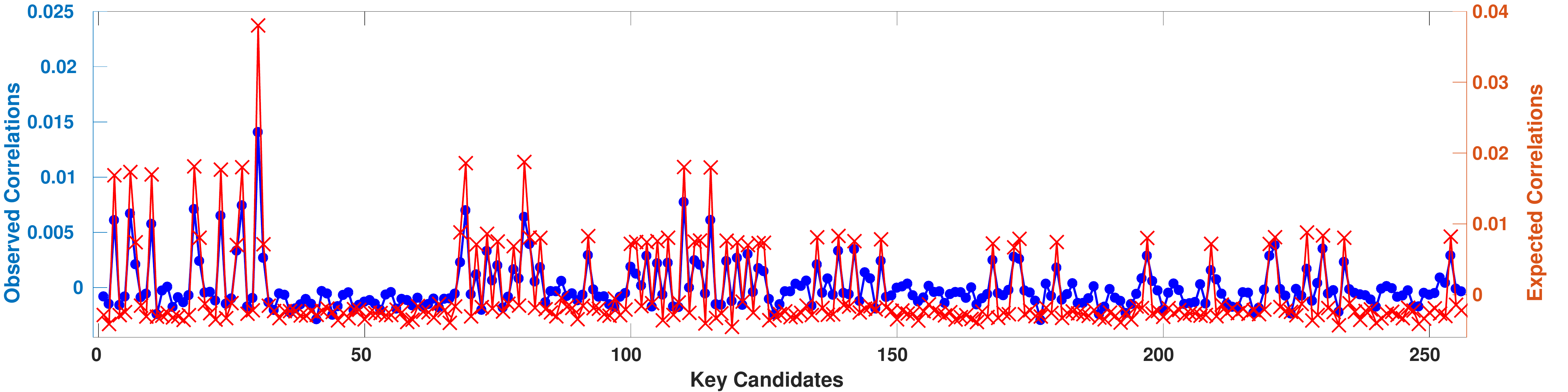}%
\label{x}}
\hfil
\caption{Linearity of the number of accesses to the first block and the execution time of AES: The synthetic correlation and \Attack\ observed correlation show similar behavior with slight difference due to the added noise.}
\label{aes_key_linearity}
\end{figure*}

\begin{figure*}[t!]
\centering
\begin{minipage}{.49\textwidth}
\includegraphics[width=.97\linewidth]{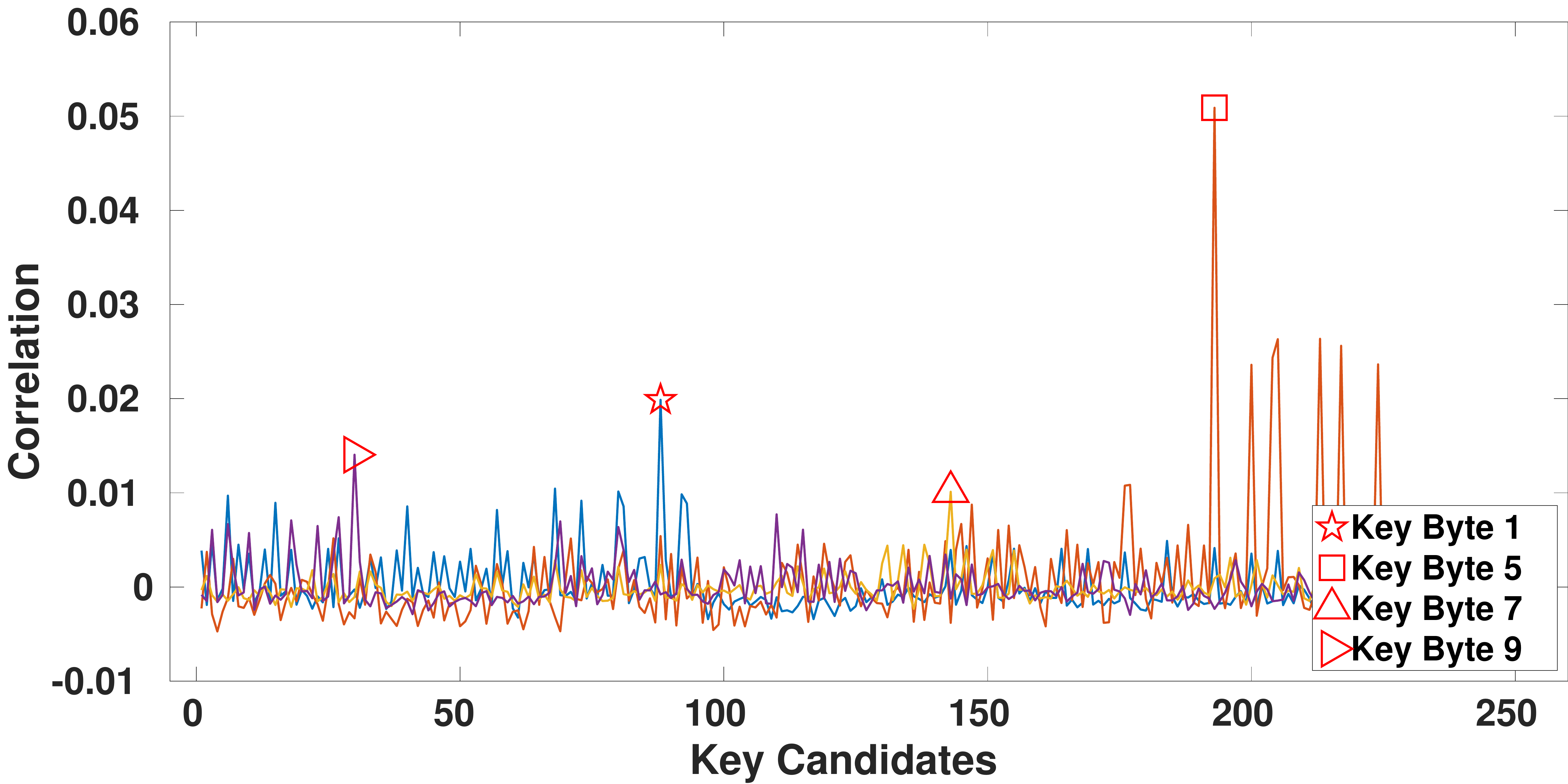}
\caption{Correlations for 4 key bytes using 2 million observations. Correct key byte candidates have the highest correlations.}
\label{fig:aes_key_cor}
\end{minipage}\hfill
\begin{minipage}{.49\textwidth}
\includegraphics[width=.97\linewidth]{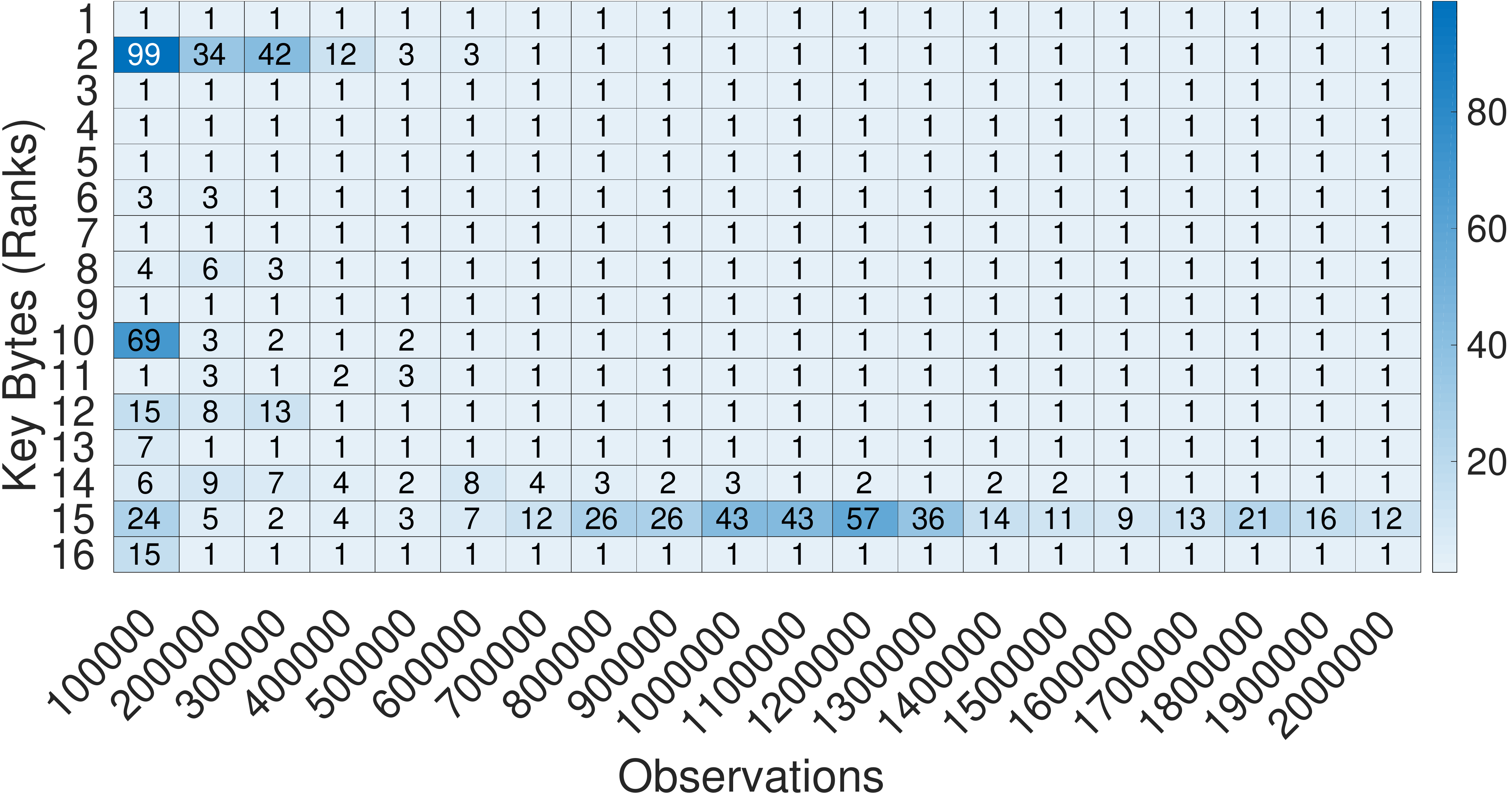}
\caption{The rank for correct key bytes are reduced with more observation. After 2 million observations, 15 out of 16 key bytes are recovered.}
\label{fig:aes_key_ranks}
\end{minipage}
\end{figure*}

Based on \Attack~4-byte granular leakage channel, and the design of AES, we can create a simple correlation model to attack this implementation. The accessed table index of the last round for a given ciphertext byte $c$ and key byte $k$ is given as $index = S^{-1}(c \oplus k)$. We define matrix $\mathbf A$ for the access profile where each row corresponds to a known ciphertext, and each column indicates the number of accesses when $index < 4$. While we assume that the attacker causes slow-downs to the first 4-byte block of S-Box, we define matrix $\mathbf L$ for leakage where each row corresponds to a known ciphertext and each column indicates the victim's encryption time. Then our correlation attack is defined as the correlation between $\mathbf A$ and $\mathbf L$, in which the higher the number of accesses, the higher the running time. Our results will verify that correlation is high, even though the implementation has dummy accesses to the monitored block. These can be ignored as noise, slightly reducing our maximum achievable correlation.

\smallskip
\noindent\textbf{AES Key Recovery Results on Synthetic Data:}
We first verified the correctness of our correlation model on synthetic data using a noise free leakage (generated by PIN~\cite{pintool}). For each of the 16 key bytes using a vector that matches exactly to the number of accesses to the targeted block of S-Box for different ciphertexts, all the correct key bytes will have the highest correlation after 32,000 observations with the best and worst correlations of 0.046 and 0.029 respectively.

\smallskip
\noindent\textbf{AES Key Recovery Results using \Attack:} Relying on the verification of Synthetic Data, we plugged in the real attack data vector, which consists of pairs of ciphertext and time measured through repeated encryption of unknown data blocks. Results on AES show that we can effectively exploit the timing information, and break the so-called constant-time implementation. The victim execution of AES encryption function takes about 1700 and 2000 cycles without and with an active thread on the logical processor pair, respectively. The target AES implementation performs 640 memory accesses to the S-Box, including dummy accesses. If the spy thread constantly writes to any address that collides with a S-Box block offset, the time will increase to a range between 2000 and 2300 cycles. The observed variation in this range has a correlation with the number of accesses to that block. Figure~\ref{aes_key_linearity} shows the linear relationship between the correlation of synthetic data and real attack data for one key byte after 2 million observations. Most of the possible key candidates for a target key byte have a matching peak and hill between the two observations. The highest correlation points in both cases declare the correct key byte (0.038 red, 0.014 blue). The quantitative difference is due to the expected noise in the real measurements.

Figure~\ref{fig:aes_key_cor} shows the correlation of 4 different key bytes after 2 million observations with the correct key bytes having the highest correlations. Our repeated experiments with different keys and ciphertexts show that 15 correct key bytes have the highest correlation ranks, and there is only the key byte at index 15 that has a high rank but not necessarily the highest. Figure~\ref{fig:aes_key_ranks} shows the key ranks over the number of observations. Key byte ranks take values between 1 and 256, where 1 means that the correct key byte is the most likely one. As it is shown, after only 200,000 observations, the key space is reduced to a computationally insecure space and a key can be found with an efficient key enumeration method~\cite{glowacz2015simpler}. After 2 million observations, all key bytes except one of them are recovered. The non-optimized implementation of this attack processes this amount of information in 5 minutes.

\begin{figure*}[t]
\centering
\includegraphics[width=.7\linewidth]{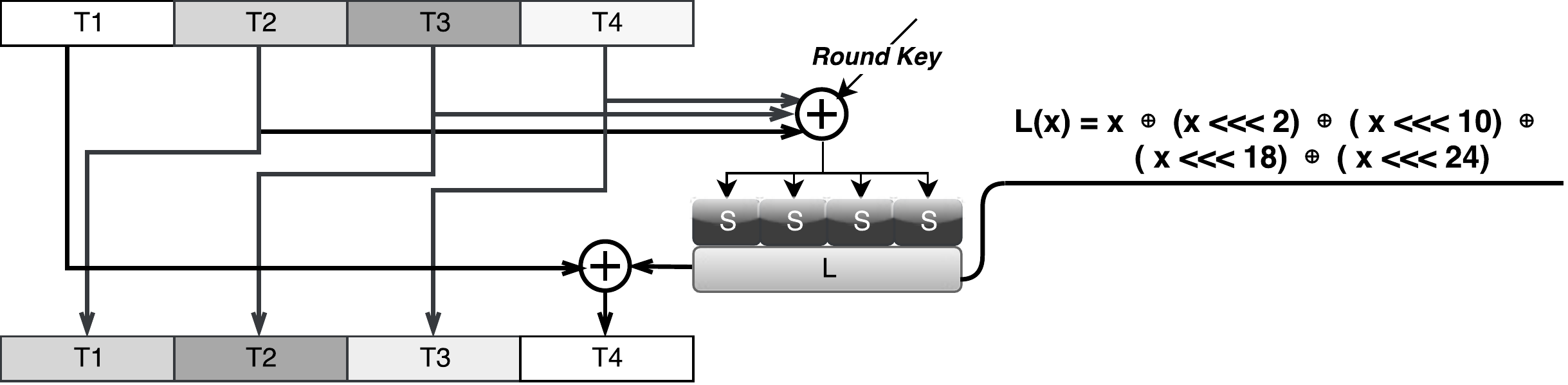}
\caption{SM4 Feistel Structure: In each round, the last three words from the state buffer and the round key will be added. Each byte of the output will be replaced by S-Box lookup. The function L performs a linear bit permutation.}
\label{fig:sm4}
\end{figure*}

\subsection{Attack 2: IPP Cache Protected SM4}
SM4 is a block cipher\footnote{Formerly SMS4, the standard cipher for Wireless LAN Wired Authentication and Privacy Infrastructure (WAPI)} that features an unbalanced Feistel structure and supports 128-bit blocks and keys~\cite{diffie2008sms4}. SM4 design is known to be secure and no relevant cryptanalytic attacks exist for the cipher. Figure~\ref{fig:sm4} shows a schematic of one round of SM4. T1-T4 are $4\times32$-bit state variables of SM4. Within each round, the last three state variables and a 32-bit round key are mixed, and each byte of the output will be replaced by a non-linear S-Box value. After the non-linear layer, the combined 32-bit output of S-Boxes $x$ are diffused using the linear function L. The output of $L$ is then mixed with the first 32-bit state variable to generate a new random 32-bit state value. The same operation is repeated for 32 rounds, and each time a new 32-bit state is generated as the next round T4 state. The current T2, T3, T4  are treated as T1, T2, and T3 for the next round. The final 16 bytes of the entire state after the last round produce the ciphertext. SM4 Key schedule produces $32\times32$-bit round keys from a 128-bit key. Since the key schedule is reversible, recovering 4 repeated round keys provides enough entropy to reproduce the cipher key.

All the SM4 operations except the S-Box lookup are performed in 32-bit word sizes. Hence, SM4 implementation is both simple and efficient on modern architectures. We chose the function \texttt{cpSMS4\_Cipher} from Intel IPP Cryptography library. Our target is based on the straightforward cipher algorithm with addition of S-Box cache state normalization. We recovered this implementation through reverse engineering of Intel IPP binaries. The implementation preloads four values from different cache lines of S-Box before the first round, and it mixes them with some dummy variables, forcing the processor to fill the relevant cache lines with S-Box table. This cache prefetching mechanism protects SM4 against asynchronous cache attacks. On our experimental setup, the implementation runs in about 700 cycles, which informs us that this implementation maintain a high speed while secure against asynchronous attacks. Interrupted attacks that leak intermediate states would not be as simple, since the interruption need to happen faster than 700 cycles. We will further discuss the difficulty of correlating any cache-granular information, even if we assume the adversary can interrupt the encryption and perform some intermediate observations.

\smallskip
\noindent\textbf{Single-round attack on SM4:}
We define $c_{1}, c_{2}, c_{3}, c_{4}$ as the four 32-bit words of a ciphertext and $k_{r}$ as the secret round key for round $r$. We recursively follow the cipher structure from the last round with our ciphertext words as inputs, and write the last 5 rounds' relations as Equation~\ref{eq:sm4}. In each round, $x^{i}_{r}$ is the S-Box index, and $i$ is the byte offset of the 32-bit word $x_{r}$. With a similar approach to the attack on AES, we define matrix $\mathbf A$ for the access profile, where each row corresponds to a known ciphertext, and each column indicates the number of accesses when $x^{i}_{r} < 4$. Then we define the matrix $\mathbf L$ for the observed timing leakage and the correlation between $\mathbf A$ and $\mathbf L$ similar to the AES attack. In contrast, S-Box indices in the AES attack are defined based on a non-linear inverse S-Box operation of key and ciphertext, which eventually maps to all possible key candidates. In SM4, the index $x^{i}_{r}$ is defined before any non-linear operation. As a result, an attack capable of distinguishing accesses of 4 out of 256 S-Box entries reveals only 6 bits per key byte. In the mentioned relations, performing the attack using this model on $x^i_{32}$, recovers the 6 most significant bits of each key byte $i$ for the last round key (Total of 24 out of the 32 bits).

\begin{small}
\begin{align}
&x_{32} = c_{1} \oplus c_{2} \oplus c_{3} \oplus k_{32}\hspace{5ex} d_{2} = c_{1}, d_{3} = c_{2}, d_{4} = c_{3}\nonumber\\
&d_{1} = L(s(x^1_{32}), s(x^2_{32}), s(x^3_{32}), s(x^4_{32})) \oplus c_{4}\nonumber\\
&x_{31} = d_{1} \oplus d_{2} \oplus d_{3} \oplus k_{31}\hspace{5ex} e_{2} = d_{1}, e_{3} = d_{2}, e_{4} = d_{3}\nonumber\\ 
&e_{1} = L(s(x^1_{31}), s(x^2_{31}), s(x^3_{31}), s(x^4_{31})) \oplus d_{4}\nonumber\\
&x_{30} = e_{1} \oplus e_{2} \oplus e_{3} \oplus k_{30}\hspace{5ex} f_{2} = e_{1}, f_{3} = e_{2}, f_{4} = e_{3}\nonumber\\
&f_{1} = L(s(x^1_{30}), s(x^2_{30}), s(x^3_{30}), s(x^4_{30})) \oplus e_{4}\nonumber\\
&x_{29} = f_{1} \oplus f_{2} \oplus f_{3} \oplus k_{29}\hspace{5ex} g_{2} = f_{1}, g_{3} = f_{2}, g_{4} = f_{3}\nonumber\\
&g_{1} = L(s(x^1_{29}), s(x^2_{29}), s(x^3_{29}), s(x^4_{29})) \oplus f_{4}\nonumber\\
&x_{28} = g_{1} \oplus g_{2} \oplus g_{3} \oplus k_{28}\nonumber\\
\label{eq:sm4}
\end{align}
\end{small}

\smallskip
\noindent\textbf{Multi-round attack on SM4:}
The relationship for round $31$ can be used not only to recover 6-bit key candidates of round $31$, but also the remaining unknown 8 bits of entropy for round $32$. This is due to the linear property of function L and the recursive nature of newly created state variables. After the attack on round $32$, similar to the round key, we only have certainty about 24 bits of the new state variable $d_{1}$, but this information will be propagated as the input to round $31$. The next round of attack for key byte of round $31$ needs more computation to process an 8 bit of unknown key and 8 bit of unknown state (total of 16 bit), but this is computationally feasible, and the 8-bit key from round $32$ with highest correlation can be recovered by attacking the S-Box indices in round $31$. We recursively applied this model to each round resulting a correlation attack with the following steps, which gives us enough entropy to recover the key:

\begin{figure*}[t!]
\centering
\begin{minipage}{.49\textwidth}
\includegraphics[width=\linewidth]{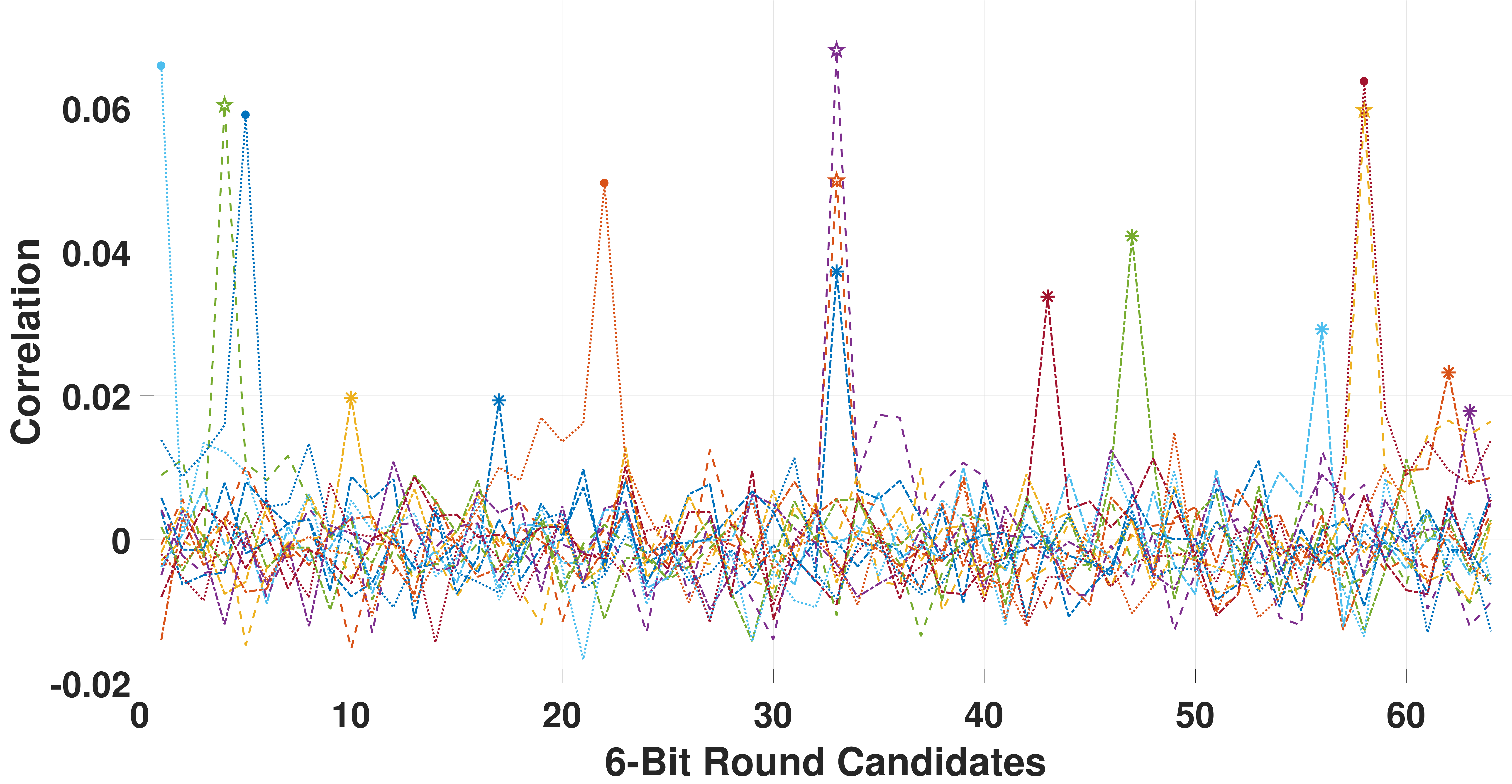}
\caption{Correlations for SM4 6-bit keys of the last 4 32-bit round key recovered through 5 rounds of attack using 40,000 observations.}
\label{fig:sm4_6bit}
\end{minipage}\hfill
\begin{minipage}{.49\textwidth}
\includegraphics[width=\linewidth]{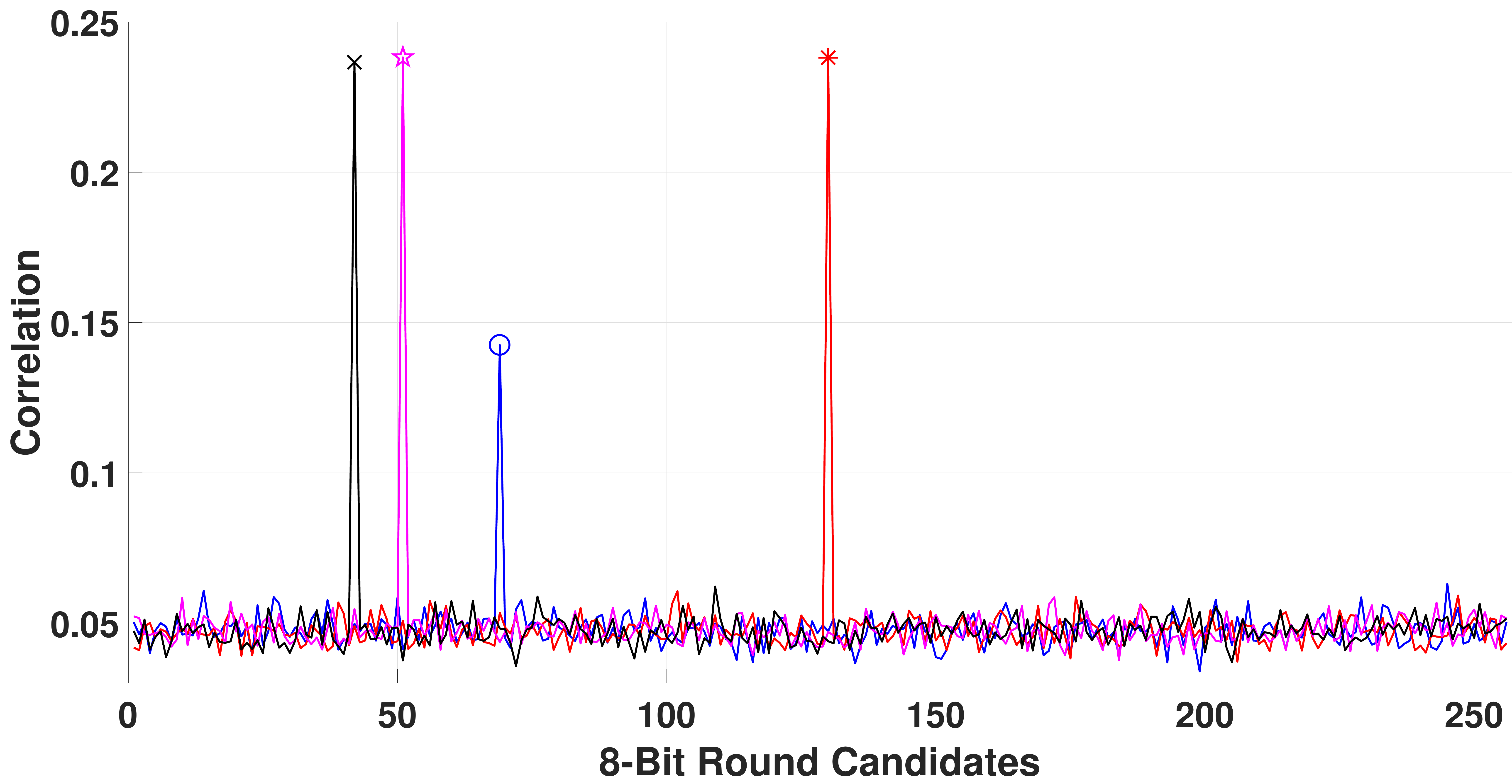}
\caption{The accumulated correlations for SM4 8-bit keys after 5 rounds using 40,000 observations. Each correct candidate has the highest correlation.}
\label{fig:sm4_8bit}
\end{minipage}
\end{figure*}

\small 
\setlength{\columnsep}{0in}
\begin{enumerate}
\item $x_{32}$ $\rightarrow$ 24 bits of $k_{32}$.
\item $x_{31}$ $\rightarrow$ 24 bits of $k_{31}$ + 8 bits of $k_{32}$
\item $x_{30}$ $\rightarrow$ 24 bits of $k_{30}$ + 8 bits of $k_{31}$
\item $x_{29}$ $\rightarrow$ 24 bits of $k_{29}$ + 8 bits of $k_{30}$
\item $x_{28}$ $\rightarrow$ 24 bits of $k_{28}$ + 8 bits of $k_{29}$
\item Recover the key from $k_{32}, k_{31}, k_{30}, k_{29}$
\end{enumerate}
\normalsize

\smallskip
\noindent\textbf{SM4 Key Recovery Results on Synthetic Data:}
Our noise-free synthetic data shows that 3000 observations are enough to find all correct 6-bit and 8-bit round key candidates with the highest correlations. Even in an interrupted cache attack or without cache protection, targeting this implementation using a cache-granular information would be much harder and inefficient due to the lack of intra cache-line resolution. If we only distinguish the 64-byte cache lines out of a 256-byte S-Box, we only learn $4\times2$-bit (total of 8 bits) out of 32-bit round keys, and on each round, we need to solve 8 bits + 24 bits of uncertainty. Although solving 32-bit of uncertainty sounds possible for a noise-free data, it is computationally much harder in a practical noisy setting. Our intra cache line leakage can exploit SM4 efficiently in a known-ciphertext scenario, while the best efficient cache attack on SM4 requires chosen plaintexts~\cite{nguyen2012improved}.

\smallskip
\noindent\textbf{SM4 Key Recovery Results using \Attack:}
The results on SM4 show even more effective key recovery against this implementation compared to AES. Figure~\ref{fig:sm4_6bit} shows the correlation for 6-bit round keys after 5 rounds of repeated attack, and the correlation for 12-bit key candidates can be seen in Figure~\ref{fig:sm4_8bit}. The attack expects assurance on the correct key candidates for each round of attack before proceeding to the next round due to the recursive structure of SM4. In our experiment using real measurement data, we have noticed that 40,000 observations are sufficient to have assurance of correct key candidates with the highest correlations. Our implementation of the attack can recover the correct 6-bit and 8-bit keys, and it takes about 5 minutes to recover the cipher key. In Figure~\ref{fig:sm4_8bit}, we plotted the accumulated per byte correlations for all 8-bit candidates within each round of attack. During the computation of 6-bit candidates, the 8-bit candidates relate to 4 different state bytes. This accumulation greatly increases the result and the correct 8-bit key candidates have a very high aggregated correlation compared to the 6-bit candidates.

\section{\Attack ing SGX Enclave} \label{sec:sgx}
Intel SGX is a trusted execution environment (TEE) extension released as part of Skylake processor generation~\cite{linuxsgx}. The main goal of SGX is to protect runtime data and computation from system and physical adversaries. Having said that, SGX must remain secure in the presence of malicious OS, thus modification of OS resources for facilitation of side-channel attacks is relevant and within the considered threat model. Previous works demonstrate high resolution attacks with 4\,kB page~\cite{xu2015controlled,van2017telling} and 64\,B cache line granularity~\cite{BrasserGrand,moghimi2017cachezoom}. Intel declared microarchitectural leakages out of scope for SGX, thus pushing the burden of writing leakage free constant-time code onto enclave developers. 
Indeed, Intel follows this design paradigm and ensures constant cache-line accesses for its AES implementation, making it resistant to \emph{all} previously known microarchitectural attacks in SGX. 

In this section, we verify that \Attack\ is also applicable to SGX enclaves, as there is no fundamental microarchitectural changes to resist against memory false dependencies. We repeat the key recovery results against Intel's constant-time AES implementation after moving it into an SGX enclave. The results verify the exploitability of intra cache level channel against SGX secure enclaves. In fact, the attack can be reproduced in a straightforward manner. The only difference is a slower key recovery due to the increased measurement noise resulting from the enclave context switch.

\begin{figure*}[t!]
\centering
\begin{minipage}{.49\textwidth}
\centering
\includegraphics[width=.98\linewidth]{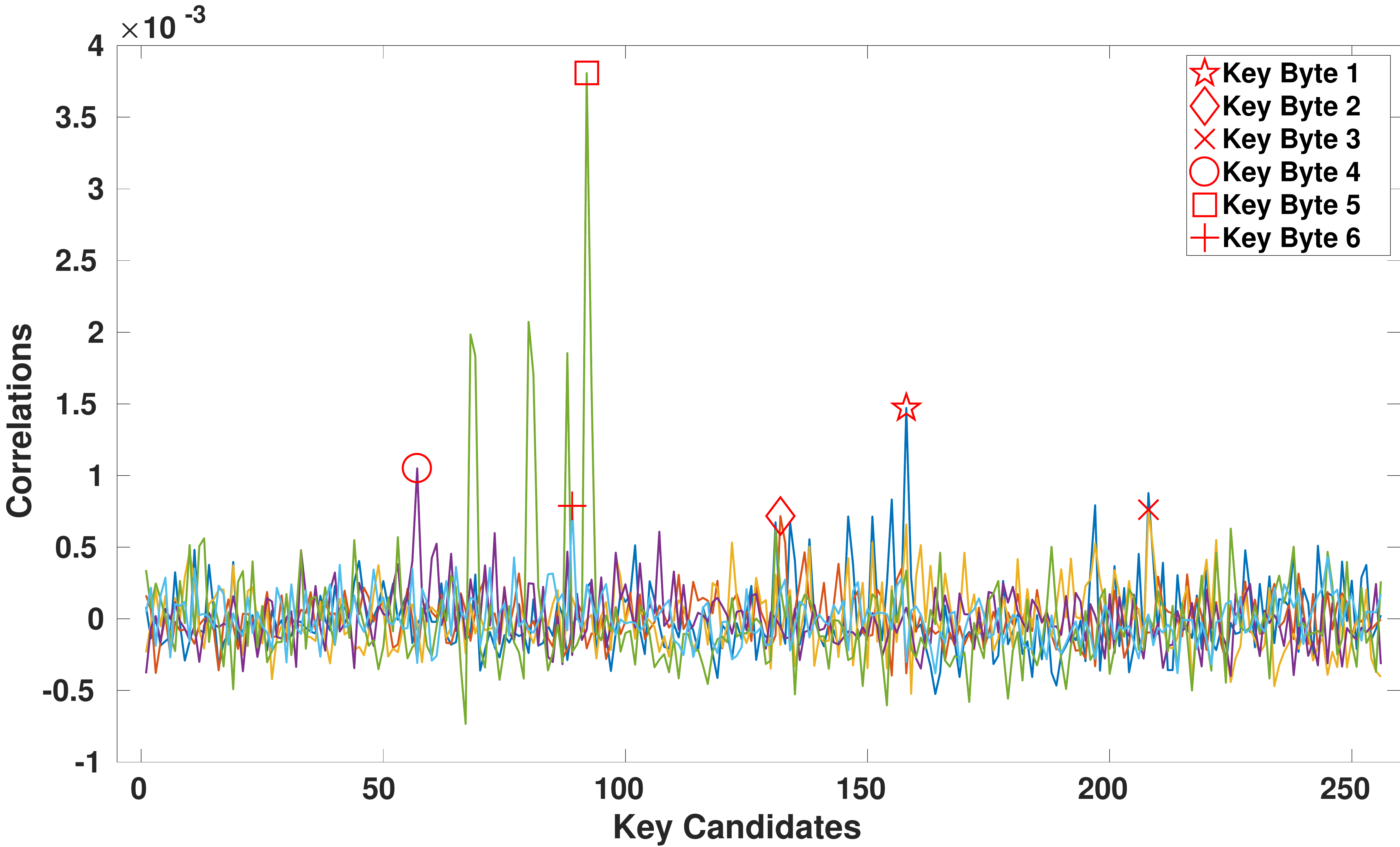}
\caption{Correlations for 6 key bytes using 5 million observations. All of the correct candidates have the highest correlations. }
\label{fig:sgx_aes_key_cor}
\end{minipage}\hfill
\begin{minipage}{.49\textwidth}
\centering
\includegraphics[width=.95\linewidth]{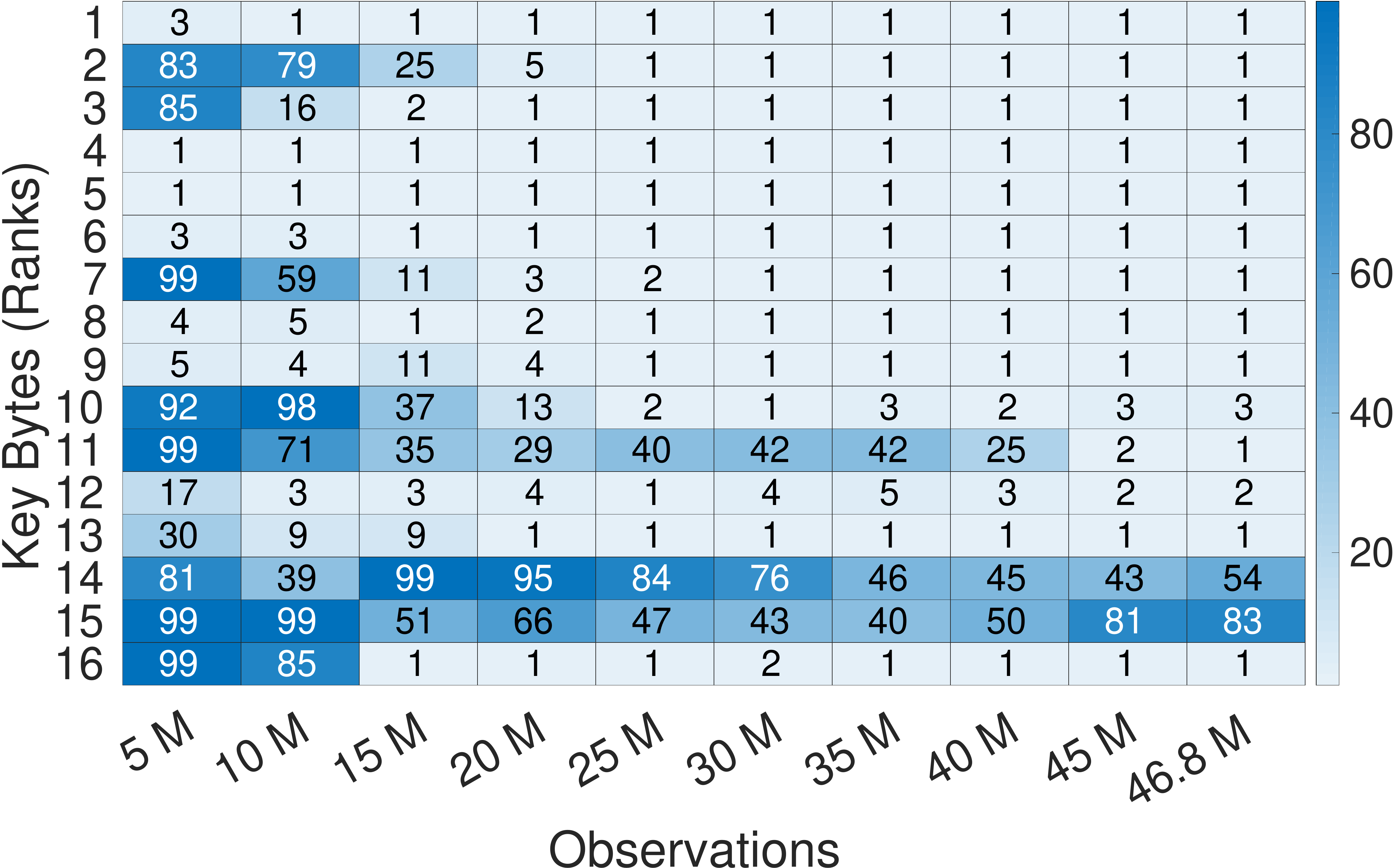}
\caption{The rank for correct key bytes with respect to the number of observations. Using the entire data set, after filtering the outliers, we can recover 14 out of 16 key bytes.}
\label{fig:sgx_aes_key_ranks}
\end{minipage}
\end{figure*}

\subsection{SGX Enclave Experimental Setup and Assumptions}
Following the threat model of \emph{CacheZoom}~\cite{lee2016inferring,moghimi2017cachezoom}, we assume that the system adversary has control over various OS resources. Please note that SGX was exactly designed to thwart the threat of such adversaries. The adversary uses its OS-level privileges to decrease the setup noise: We isolate one of the physical cores from the rest of the running tasks, and dedicate its logical processors to~\Attack~write conflict thread and the victim enclave. We further disable all the non-maskable interrupts on the target physical core and configure the CPU power and frequency scaling to maintain a constant frequency. We assume that the adversary can measure the execution time of an enclave interface that performs encryption, and the enclave interface only returns the ciphertext to the insecure environment. Both plaintexts and the secret encryption key are generated at runtime using \emph{RDRAND} instruction, and they never leave the secure runtime environment of SGX enclave. The \emph{RDTSC} instruction cannot be used inside an enclave. The attacker uses it right before the call to the enclave interface and again right after the enclave exit.  As a result, the entire execution of the enclave interface, including the AES encryption, is measured. As before, an active thread causing read-after-write conflicts to the first 4-byte of AES S-Box is executed on the neighboring virtual processor of the SGX thread. 

\subsection{AES Key Recovery Results on SGX}
Execution of the same AES encryption function as Section~\ref{sec:aes} inside an SGX enclave interface takes an average of 14,600 cycles with an active thread causing read-after-write conflicts to the first 4-byte of AES S-Box. The additional overhead is caused by the enclave context switch, which significantly increases the noise of the timing channel due to the variable timing behavior. Having that, this experiment shows a more practical timing behavior where adversaries cannot time the exact encryption operation, and they have to measure the time for a batch of operations. This not only shows that SGX is vulnerable to~\Attack~attack, but it also demonstrates that \Attack\ is applicable in a realistic scenario. Figure~\ref{fig:sgx_aes_key_cor} shows the key correlation results using 50 million timed encryptions in SGX, collected in 10 different time frames. We filtered outliers, i.e. measurements with high noise by only considering samples that are in the range of 2000 cycles of the mean. Among the 50 million samples, 93\% pass the filtering, and we only calculated the correlations for the remaining traces. Figure~\ref{fig:sgx_aes_key_ranks} shows that we can successfully recover 14 out of 16 key bytes, revealing sufficient information for key recovery after 20 million observations. 

These results show that even cryptographic libraries designed by experts that are fully aware of current attacks and of the leakage behavior of the target device may fail at writing unexploitable code. Modern microarchitectures are so complex that assumptions such as \emph{constant cache line profiles} result in unexploitable constant-time implementations are seemingly impossible to fulfill. 

\section{Discussion}

The \texttt{Safe2Encrypt\_RIJ128} AES implementation has been designed to achieve a constant cache access profile by ensuring that the same cache lines are accessed every time regardless of the processed data.  The 4-byte spatial resolution of \Attack , however, thwarts this countermeasure by providing intra cache-line resolution. One approach to restore security and protect against \Attack is to apply constant memory accesses with a 4-byte granularity. That would require accessing every fourth  byte of the table for each memory lookup for the purpose of maintaining a uniform memory footprint. At that point, it might be easier to just do a \emph{true} constant time implementation and access \emph{all} entries each time, resting assured that there is no other effect somewhere hidden in the microarchitecture resulting in a leak with byte granularity. As we discussed in the related work, system-wide defense proposals that apply to cache attacks are not relevant and cannot detect or prevent~\Attack. 
Also, an adversary performing the \Attack\ attack does not need to know about the offset of S-Box in the binary, since she can simply scan the 10-bits address entropy through introducing conflicts to different offsets and measuring the timing of victim. This is important when it comes to obfuscated binaries or scenarios, where the offset of S-Box is unknown.

Hardware based, e.g, AES-NI or hardware assisted, e.g, SIMD-based bit-sliced implementations of AES or SM4 should exclusively be used to protect the targeted implementation in an efficient manner. Intel IPP has different variants optimized for various generations of Intel instruction sets~\cite{intelcpudispatch}. Intel IPP features different implementations of AES as well as SM4 in these variants. A list of these variants and implementations are given in Table~\ref{tab:implementations}. All of them have at least one vulnerable implementation. In cases where there is an implementation based on the AES-NI instruction set (or SSSE3 respectively), the library falls back to the basic version at runtime if the instruction set extensions are not available. The usability of this depends on the compilation and runtime configuration. Developers are allowed to statically link to a more risky variants~\cite{intelIPPlinkage}, and they need to assure not to use the vulnerable versions during linking. These ciphers should be avoided in cases where the hardware does not provide support, e.g, Core and Nehalem does not support AES-NI, e.g, AES-NI can be disabled in some BIOS. After all, the current hardware support for cryptographic primitives are restricted and if any other cipher is demanded, this limitation and vulnerability endangers the security of cryptographic systems. A temporary workaround to defend against the source of leakage on current Intel microarchitectures is to disable hyper-threading.
 
Prior to \Attack\, it might have seemed reasonable to design SGX enclaves under the paradigm that constant cache line accesses result in leakage-free code. However, the increased 4-byte intra cache-line granularity of \Attack\ shows that only code with true constant-time properties, i.e. constant execution flow and constant memory accesses can be expected to have no remaining leakage on modern microarchitectures. 

\smallskip
\noindent\textbf{Responsible Disclosure} 
We have informed the Intel Product Security Incident Response Team of our findings on August 2nd, 2017. They have acknowledged the receipt and are currently reviewing our findings. 

\begin{table}[t!]
\begin{center}
    \captionof{table}{SM4 and AES implementations in all variants of Intel IPP library version 2017 update 3~\cite{intelcpudispatch}. The variants will be merged at linker and each variant is optimized for a different generation of the Intel instruction set~\cite{intelIPPlinkage}. Developers can statically link specific variants with single processor static linking mode~\cite{intelcpudispatch}.} \label{tab:implementations} 
    \begin{tabular}{ | p{1.8cm} | l | p{.3cm} | p{.3cm} | p{.3cm} | p{.9cm} | }
    \hline
     \textbf{Implementation Technique} & \textbf{Function Name} & \textbf{l9 n0 y8 k0 e9} & \textbf{m7 mx} & \textbf{n8} & \textbf{Linux SGX SDK}\\ \hline
     \small{AES-NI} & \scriptsize{Encrypt\_RIJ128\_AES\_NI}  & $\checkmark$ & $\times$  & $\times$ & $\checkmark$ (prebuilt)\\ \hline
     \small{AES} \scriptsize{Bitsliced} & \scriptsize{SafeEncrypt\_RIJ128} & $\checkmark$ & $\times$ & $\checkmark$ & $\checkmark$ (prebuilt)\\ \hline
     \small{AES} \scriptsize{Constant-Time} & \scriptsize{Safe2Encrypt\_RIJ128} & $\times$ & $\checkmark$  & $\times$ & $\checkmark$ (source)\\ \hline
     \small{SM4} \scriptsize{Bitsliced using AES-NI} & \scriptsize{cpSMS4\_ECB\_aesni} & $\checkmark$ & $\times$ & $\times$ & N/A\\ \hline
     \small{SM4} \scriptsize{Cache Normalization} & \scriptsize{cpSMS4\_Cipher} & $\checkmark$ & $\checkmark$ &  $\checkmark$ & N/A\\ \hline
    \end{tabular}
\end{center}
\end{table}

\begin{table}[t!]
    \caption{Intel processor families and availability of the leakage channels. Major Intel processors suffer from 4k aliasing, and are vulnerable to \Attack~\cite{agneroptimize}.} \label{tab:cpus}
    \centering
    \begin{tabular}{ | l | l | p{1.4cm} | p{1.4cm} |}
    \hline
     \textbf{Release} & \textbf{Family} & \textbf{Cache Bank Conflicts} & \textbf{4K Aliasing}\\ \hline
     \scriptsize{2006} & \scriptsize{Core}  & $\checkmark$ & $\checkmark$ \\ \hline
     \scriptsize{2008} & \scriptsize{Nehalem}  & $\times$ & $\checkmark$ \\ \hline
     \scriptsize{2011} & \scriptsize{Sandy bridge}  & $\checkmark$ & $\checkmark$ \\ \hline
     \scriptsize{2013} & \scriptsize{Silvermont, Haswell, Broadwell} & $\times$ & $\checkmark$ \\ \hline
     \scriptsize{2015} & \scriptsize{Skylake}  & $\times$ & $\checkmark$ \\ \hline
     \scriptsize{2016} & \scriptsize{KabyLake}  & $\times$ & $\checkmark$ \\ \hline
    \end{tabular}
\end{table}

\section{Conclusion}
This work proposes \Attack, a new side-channel attack based on false dependencies. For the first time, we discovered new aspects of this side channel and its capabilities, and show how to extract secrets from modern cryptographic implementations. \Attack\ uses false read-after-write dependencies to slow down accesses of the victim to a particular 4-byte memory blocks \emph{within} a cache line. The resulting latency of otherwise constant-time implementations was exploited with state-of-the art timing side-channel analysis techniques. We showed how to apply the attack to two recent implementations of AES and SM4. According to the available resources, the source of leakage for\Attack~attack is present in all Intel CPU families released in the last 10 years~\cite{agneroptimize,inteloptimze}. Our results also verified that \Attack~is the first intra cache level attack applicable to SGX enclaves. Table~\ref{tab:cpus} highlights the availability of the cache bank conflicts and 4k aliasing leakage source. \Attack\ is another piece of evidence that modern microarchitectures are too complex and constant-time implementations cannot simply be trusted with wrong assumptions about the underlying system. The remaining data-dependent addressing within a cache line is exploitable.

\section*{Acknowledgements}
This work is supported by the National Science Foundation, under grant CNS-1618837.

\bibliographystyle{IEEEtran}
\bibliography{IEEEabrv,reference}

\end{document}